\documentclass[twocolumn]{aastex631}

\usepackage{amsmath}
\usepackage{booktabs}

\usepackage{siunitx}
\usepackage[caption=false,font=footnotesize]{subfig}
\usepackage{tikz}
\usetikzlibrary{arrows.meta,positioning,calc,shapes.geometric}
\graphicspath{{figures_original/}{figures_generated/}}
\newcommand{\IRR}{\mathrm{IRR}}

\makeatletter
\long\def\frontmatter@title@above{}
\makeatother

\begin{document}

\title{A Frequency-Agile Image-Reject Channelizer for Spaceborne 0--50 MHz Radio Spectroscopy}
\shorttitle{Spaceborne Low-Frequency Radio Channelizer}
\shortauthors{ElSayed}

\correspondingauthor{Hassan ElSayed}
\email{hassan.elsayed@wisc.edu}
\author{Hassan ElSayed}
\affiliation{Universit\'e Pierre et Marie Curie (UPMC), 4 place Jussieu, 75252 Paris Cedex 05, France}
\affiliation{Centre National de la Recherche Scientifique (CNRS), Observatoire de Paris, 5 place Jules Janssen, F-92195 Meudon Cedex, France}

\begin{abstract}
Spaceborne low-frequency radio instruments open spectral windows that are inaccessible or strongly compromised from the ground and are used for measurements ranging from solar radio bursts and plasma waves to the low-frequency Galactic background.  We investigate a frequency-agile analog receiver intended as the channelization stage of a compact spaceborne radio spectrometer.  The receiver covers \SIrange{0}{50}{MHz}, selects a \SI{3}{MHz} RF sub-band, and translates it to \SIrange{0}{3}{MHz} with a design goal of more than \SI{40}{dB} image rejection.  A behavioral trade study compares Hartley, Weaver, and double-conversion double-quadrature architectures under I/Q phase and gain mismatch.  The Weaver topology is selected because it retains the required rejection when the tunable first local oscillator is driven by quadrature square waves.  Pairing the lower and upper halves of the observing band as image partners reduces the required first-LO tuning span from \SI{50}{MHz} to \SI{25}{MHz}, while a second LO remains fixed at \SI{26}{MHz}.  A schematic-level CMOS realization using a modified Gilbert-cell mixer, fourth-order active filters, and a compensated operational amplifier is evaluated in Cadence/Spectre.  End-to-end simulations give \SI{42.9}{dB} image rejection at \SI{1}{\degree} quadrature error and \SI{43.7}{dB} in the all-worst-speed process corner; the \SI{40}{dB} threshold is crossed near \SI{58}{\celsius}.  The results establish the frequency plan and schematic feasibility of the analog channelizer, while identifying I/Q matching and square-wave-LO harmonics as the dominant limitations.  Noise, linearity, absolute calibration, sweep cadence, post-layout effects, radiation response, and measured sensitivity remain to be established for a complete flight instrument.
\end{abstract}


\section{Introduction}
Low-frequency radio observations probe physical regimes that are difficult or impossible to access with conventional ground-based facilities.  At the lowest frequencies, ionospheric propagation and terrestrial radio-frequency interference restrict the observable sky, motivating radio instruments above the ionosphere and, ultimately, on the lunar far side.  Space measurements have already been used to constrain the diffuse radio sky below \SI{6}{MHz}, while current instrument concepts target broad bands extending to \SI{40}{MHz}--\SI{50}{MHz} for heliophysics, cosmology, Galactic science, and searches for planetary or stellar magnetospheric emission~\cite{bassett2023,farview2026}.  A recent example particularly close to the frequency range considered here is RAISE, a proposed compact \SI{50}{kHz}--\SI{50}{MHz} payload for solar-radio and RFI measurements from low Earth orbit~\cite{raise2026}.

Flight heritage demonstrates the scientific value of wideband radio and plasma-wave spectroscopy.  STEREO/WAVES combines multiple receivers spanning the plasma-wave and solar-radio bands through approximately \SI{16}{MHz}, together with a fixed-frequency channel near \SI{50}{MHz}~\cite{bougeret2008}.  Parker Solar Probe/FIELDS measures plasma waves and radio emission over bands extending to the tens-of-megahertz regime~\cite{bale2016}, and Solar Orbiter/RPW covers electric-field measurements from quasi-dc through \SI{16}{MHz}~\cite{maksimovic2020}.  Across these missions, receiver design is inseparable from spacecraft constraints on mass, power, telemetry, electromagnetic cleanliness, thermal operation, and calibration stability.

The present work addresses one component of such an instrument: a frequency-agile analog channelizer placed between a broadband antenna/preamplifier and a digital spectrometer.  Rather than digitizing the entire \SIrange{0}{50}{MHz} input band continuously, the channelizer selects a \SI{3}{MHz} spectral slice and translates it to baseband.  This can reduce the instantaneous ADC and digital-processing bandwidth, but it places stringent requirements on image rejection and local-oscillator generation.  The engineering problem is therefore directly coupled to an observing-strategy problem: analog tuning can save digital resources only if the resulting dwell time and sweep cadence remain compatible with the targeted radio phenomena.

A conventional superheterodyne receiver rejects the frequency-conversion image with a band-pass filter ahead of the mixer.  At low intermediate frequency, however, the desired channel and its image are separated by only twice the IF, making high rejection difficult without a selective RF filter.  This is particularly unattractive for compact integrated payloads.  Image-reject architectures replace part of that filter selectivity with coherent I/Q processing.  The Weaver method is well suited to a frequency-agile implementation because it suppresses an unwanted sideband through two frequency translations without the broadband \SI{90}{\degree} signal-path phase network required by a Hartley receiver~\cite{weaver1956,behbahani2001}.

This paper develops that architecture as an astrophysical-instrument front end rather than as a stand-alone RFIC benchmark.  We first define the observing-band and channelization requirements, then compare candidate image-reject topologies under phase and gain mismatch.  We show that square-wave quadrature drive is compatible with the Weaver topology, introduce an image-paired frequency plan that halves the variable-LO tuning span, and evaluate the available schematic-level circuit implementation end to end.  The final discussion translates the circuit results back into instrument terms: usable IRR margin, downstream bandwidth, calibration requirements, sweep-cadence limitations, and the measurements still needed before a flight receiver can be claimed.

\section{Instrument Concept and Frequency-Plan Requirements}\label{sec:instrument}
Figure~\ref{fig:instrument-chain} places the receiver in a generic low-frequency radio measurement chain.  A broadband antenna and preamplifier provide the RF input.  The channelizer studied here selects one \SI{3}{MHz} slice from the \SIrange{0}{50}{MHz} observing band, suppresses the conversion image, and produces a \SIrange{0}{3}{MHz} output for an ADC and digital spectrometer.  The antenna/preamplifier, digitizer, spectral estimator, telemetry, and absolute calibration system are outside the scope of the present design.

\begin{figure*}[!t]
\centering
\begin{tikzpicture}[>=Latex,font=\small]
\node[draw,rounded corners,minimum width=2.35cm,minimum height=0.90cm,align=center] (ant) at (0,0) {antenna +\\preamplifier};
\node[draw,rounded corners,fill=gray!15,minimum width=3.25cm,minimum height=1.05cm,align=center] (chan) at (4.1,0) {frequency-agile Weaver\\analog channelizer\\\SIrange{0}{50}{MHz} $\rightarrow$ \SIrange{0}{3}{MHz}};
\node[draw,rounded corners,minimum width=2.35cm,minimum height=0.90cm,align=center] (adc) at (8.25,0) {ADC + digital\\spectrometer};
\node[draw,rounded corners,minimum width=2.35cm,minimum height=0.90cm,align=center] (data) at (11.75,0) {dynamic spectra\\science data};
\draw[->,line width=0.8pt] (ant.east)--(chan.west);
\draw[->,line width=0.8pt] (chan.east)--(adc.west);
\draw[->,line width=0.8pt] (adc.east)--(data.west);
\node[below=0.72cm of chan,font=\scriptsize] {scope of this work};
\end{tikzpicture}
\caption{Instrument-level placement of the proposed receiver.  The design acts as a tunable analog channelizer: it selects a \SI{3}{MHz} RF sub-band from a \SIrange{0}{50}{MHz} input and presents that reduced bandwidth to a downstream digitizer and spectrometer.}
\label{fig:instrument-chain}
\end{figure*}

\begin{table*}[!t]
\caption{Instrument and Frequency-Plan Requirements}
\label{tab:specs}
\centering
\begin{tabular}{@{}lll@{}}
\toprule
Quantity & Requirement / design choice & Instrument role \\
\midrule
RF observing band & \SIrange{0}{50}{MHz} & wide low-frequency spectral coverage \\
Instantaneous selected bandwidth & \SI{3}{MHz} & analog sub-band presented to digitizer \\
Output band & \SIrange{0}{3}{MHz} & baseband/low-frequency ADC input \\
Image-rejection target & $>\SI{40}{dB}$ & suppress opposite-sideband contamination \\
Variable first-LO span & \SI{25}{MHz} & reduced by image-paired band partitioning \\
Second LO & fixed at \SI{26}{MHz} & translates \SIrange{23}{26}{MHz} IF to baseband \\
\bottomrule
\end{tabular}
\end{table*}

The \SI{3}{MHz} output should be understood as an instantaneous analog bandwidth, not as the spectral resolution of the final science product.  Spectral resolution would be set by the downstream digital spectrometer.  Likewise, the architecture does not specify a sweep cadence: a stepped survey of the full \SIrange{0}{50}{MHz} band requires LO settling and dwell times that must be chosen for the transient timescales of the intended observations.  These instrument-level consequences are revisited in Section~\ref{sec:discussion}.

\section{Image-Reject Receiver Architecture}\label{sec:architecture}
\subsection{Image-frequency problem}
Let the desired RF input be
\begin{equation}
    v_{\mathrm{RF}}(t)=A\cos(\omega_{\mathrm{RF}}t)
\end{equation}
and the local oscillator be $v_{\mathrm{LO}}(t)=\cos(\omega_{\mathrm{LO}}t)$. An ideal multiplier produces
\begin{equation}
\begin{split}
 v_{\mathrm{RF}}v_{\mathrm{LO}}=\frac{A}{2}\bigl[&\cos\bigl((\omega_{\mathrm{RF}}-\omega_{\mathrm{LO}})t\bigr)\\
 &+\cos\bigl((\omega_{\mathrm{RF}}+\omega_{\mathrm{LO}})t\bigr)\bigr].
\end{split}
\label{eq:mixing}
\end{equation}
After low-pass filtering, the difference-frequency component is retained. If the desired signal satisfies $\omega_{\mathrm{RF}}=\omega_{\mathrm{LO}}-\omega_{\mathrm{IF}}$, an unwanted signal at $\omega_{\mathrm{im}}=\omega_{\mathrm{LO}}+\omega_{\mathrm{IF}}$ is translated to the same intermediate frequency. Once both tones occupy the same IF, ordinary filtering cannot separate them; either the image must be removed before mixing or the two quadrature paths must cancel it coherently~\cite{behbahani2001,lee2004}. The image-rejection ratio is defined as
\begin{equation}
    \IRR_{\mathrm{dB}}=10\log_{10}\left(\frac{P_{\mathrm{desired}}}{P_{\mathrm{image}}}\right).
\end{equation}

\subsection{Candidate topologies}
\begin{figure*}[!t]
\centering
\subfloat[Superheterodyne receiver.]{\includegraphics[width=0.29\textwidth]{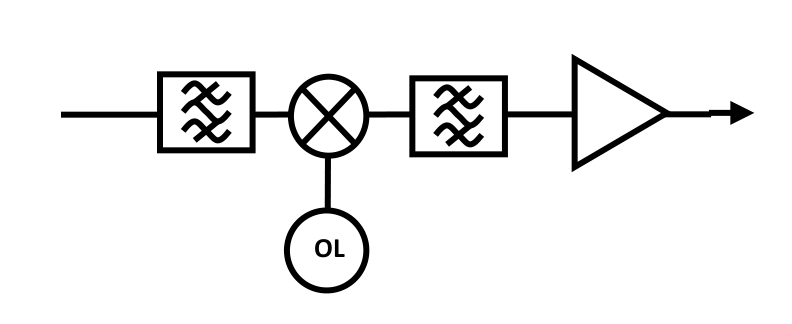}}\hfill
\subfloat[Zero-IF receiver.]{\includegraphics[width=0.19\textwidth]{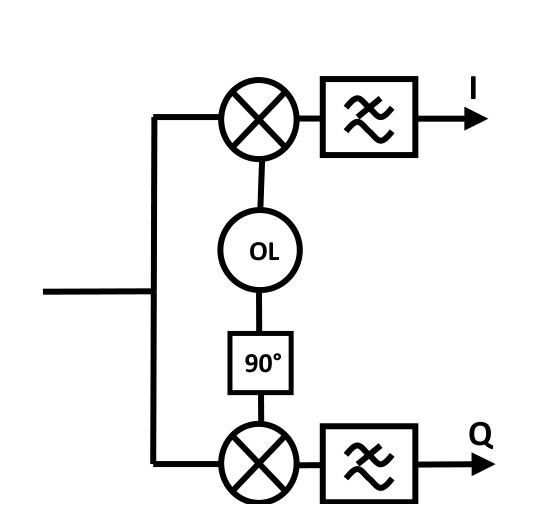}}\hfill
\subfloat[Hartley image-reject receiver.]{\includegraphics[width=0.25\textwidth]{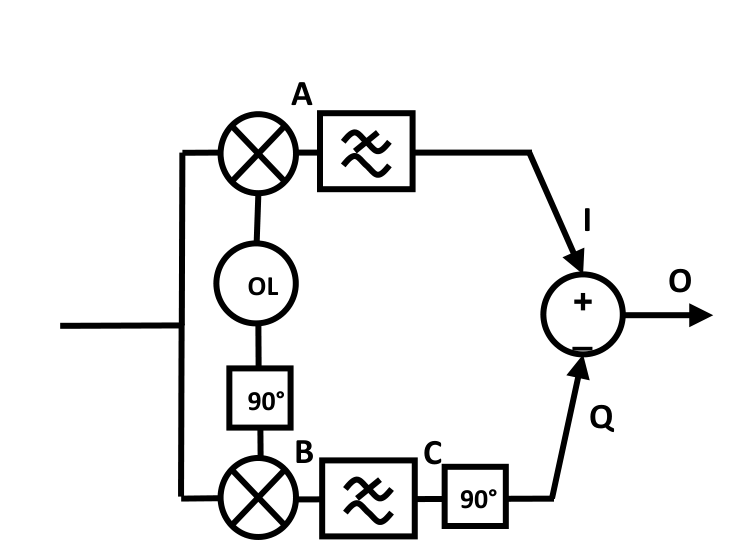}}\\[6pt]
\subfloat[Weaver image-reject receiver.]{\includegraphics[width=0.28\textwidth]{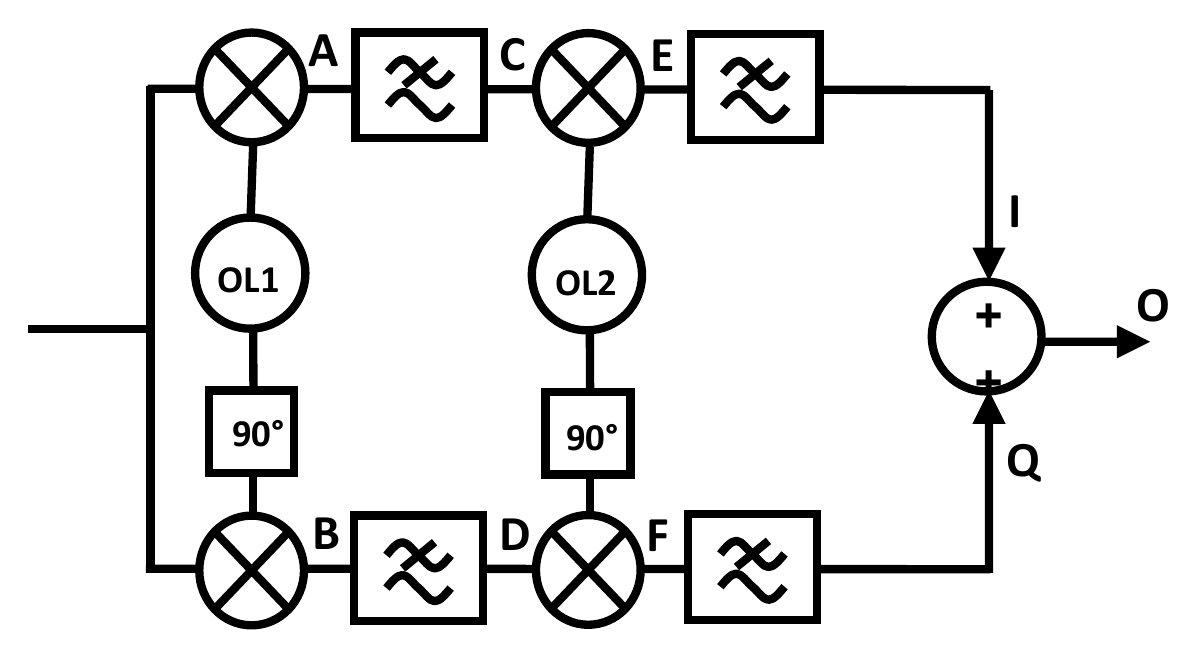}}\hfill
\subfloat[Double-conversion double-quadrature receiver.]{\includegraphics[width=0.34\textwidth]{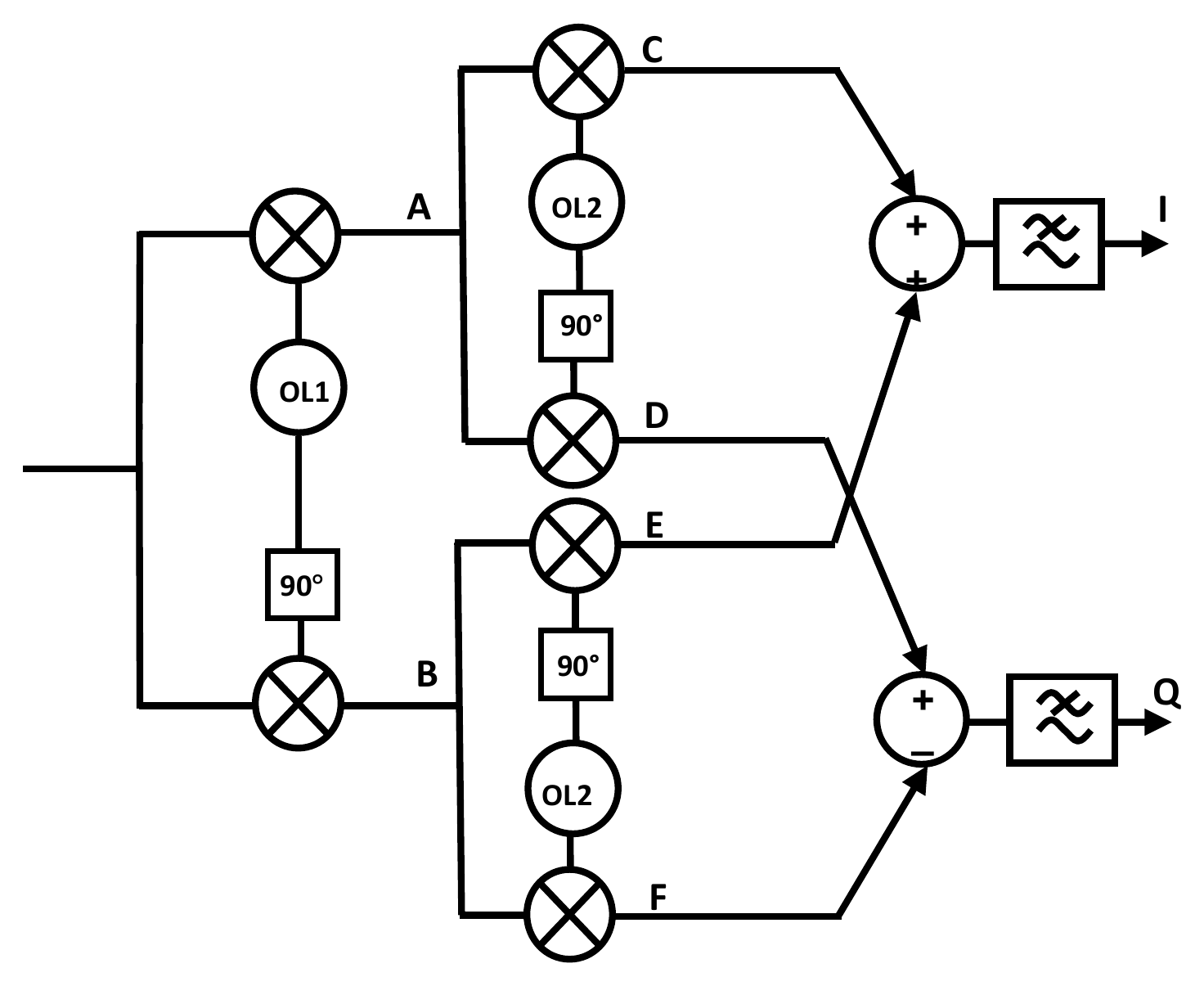}}
\caption{Receiver architectures considered in the design study.}
\label{fig:candidate-topologies}
\end{figure*}

Figure~\ref{fig:candidate-topologies} collects the principal architectures considered in the study. The superheterodyne architecture suppresses the image with an RF band-pass filter. The zero-IF architecture translates directly to dc using I/Q channels. Hartley and Weaver receivers instead arrange the desired and image components with opposite relative signs in the two branches and cancel one sideband by summation or subtraction; these architectures are classical examples of replacing filter selectivity by amplitude and phase accuracy~\cite{weaver1956,razavi1997direct,behbahani2001}. The double-conversion double-quadrature receiver uses quadrature processing at both conversion stages.

The Hartley receiver uses one quadrature LO, two low-pass filters, and a nominal \SI{90}{\degree} phase shift at the intermediate frequency. Its principal difficulty is that the IF phase shifter must maintain a nearly constant \SI{90}{\degree} offset over the tuned band. The Weaver receiver replaces the broadband IF phase shifter with a second quadrature mixing stage. This increases circuit count but makes the phase relationships easier to generate electronically. The double-conversion double-quadrature architecture is more complex still and did not provide a mismatch advantage in the behavioral study.

For a quadrature image-reject receiver with amplitude ratio $g$ between the I and Q paths and phase error $\phi$ relative to ideal quadrature, the idealized image-rejection ratio is
\begin{equation}
 \IRR=\frac{1+g^2+2g\cos\phi}{1+g^2-2g\cos\phi}.
 \label{eq:irr-mismatch}
\end{equation}
At $g=1$, (\ref{eq:irr-mismatch}) reduces to $\cot^2(\phi/2)$; at $\phi=0$, it reduces to $[(1+g)/(1-g)]^2$. These relations explain the rapid degradation caused by small quadrature errors and agree closely with the Weaver behavioral simulations. They also make clear why image rejection much above roughly \SIrange{30}{40}{dB} is difficult to maintain in untrimmed analog paths: the required phase and amplitude matching rapidly approaches sub-degree and sub-percent levels~\cite{razavi1997direct,behbahani2001}.

\subsection{Calibration alternatives}
Calibration-based architectures were considered as an alternative to relying solely on open-loop matching. This is a well-established response to the mismatch sensitivity of quadrature receivers. Der and Razavi demonstrated a CMOS image-reject receiver using sign-sign LMS adaptation to correct both phase and gain error~\cite{der2003}, while adaptive decorrelation has likewise been demonstrated in low-IF CMOS receivers~\cite{elahi2006}. More recent wideband down-converters continue to use explicit I/Q calibration when high rejection must be maintained across a large IF range; a 2023 SATCOM receiver, for example, reported \SIrange{48}{55}{dB} image rejection using automatic calibration~\cite{ma2023}. Such methods can recover high rejection, but they add detectors or auxiliary mixers, tunable delay/gain elements, control loops, calibration time, and power. Because the present receiver was intended to meet a more modest \SI{40}{dB} target over the expected mismatch range, the simpler open-loop architecture was retained.

\begin{figure*}[!t]
\centering
\subfloat[Phase-error detector.]{%
\begin{tikzpicture}[>=Latex,font=\scriptsize,baseline=(current bounding box.center)]
\node[draw,minimum width=1.2cm,minimum height=0.55cm] (iq) {I/Q};
\node[draw,right=0.45cm of iq,minimum width=1.4cm,minimum height=0.55cm] (det) {phase detector};
\node[draw,right=0.45cm of det,minimum width=1.0cm,minimum height=0.55cm] (int) {integrator};
\draw[->] (iq)--(det); \draw[->] (det)--(int);
\end{tikzpicture}}\hfill
\subfloat[Calibration path with tunable correction.]{%
\begin{tikzpicture}[>=Latex,font=\scriptsize,baseline=(current bounding box.center)]
\node[draw,minimum width=1.1cm,minimum height=0.55cm] (rx) {I/Q RX};
\node[draw,right=0.42cm of rx,minimum width=1.2cm,minimum height=0.55cm] (adj) {gain/delay};
\node[draw,right=0.42cm of adj,minimum width=0.8cm,minimum height=0.55cm] (out) {out};
\node[draw,below=0.45cm of adj,minimum width=1.25cm,minimum height=0.48cm] (cal) {calibration};
\draw[->] (rx)--(adj); \draw[->] (adj)--(out); \draw[->] (cal)--(adj);
\end{tikzpicture}}\\[8pt]
\subfloat[LMS adaptation loop.]{%
\begin{tikzpicture}[>=Latex,font=\scriptsize,baseline=(current bounding box.center)]
\node[draw,minimum width=1.1cm,minimum height=0.55cm] (rx) {I/Q RX};
\node[draw,right=0.45cm of rx,minimum width=1.0cm,minimum height=0.55cm] (err) {error};
\node[draw,below=0.45cm of err,minimum width=1.1cm,minimum height=0.5cm] (lms) {LMS update};
\draw[->] (rx)--(err); \draw[->] (err)--(lms); \draw[->] (lms.west) -| (rx.south);
\end{tikzpicture}}\hfill
\subfloat[Stored correction after calibration.]{%
\begin{tikzpicture}[>=Latex,font=\scriptsize,baseline=(current bounding box.center)]
\node[draw,minimum width=1.0cm,minimum height=0.55cm] (rx) {I/Q RX};
\node[draw,right=0.42cm of rx,minimum width=1.2cm,minimum height=0.55cm] (corr) {correction};
\node[draw,below=0.45cm of corr,minimum width=1.0cm,minimum height=0.48cm] (mem) {stored code};
\node[draw,right=0.42cm of corr,minimum width=0.8cm,minimum height=0.55cm] (out) {out};
\draw[->] (rx)--(corr); \draw[->] (mem)--(corr); \draw[->] (corr)--(out);
\end{tikzpicture}}
\caption{Conceptual calibration strategies for correcting I/Q mismatch. These original block diagrams summarize the classes of approaches discussed in~\cite{der2003,elahi2006,ma2023}; calibration was not adopted in the present receiver.}
\label{fig:calibration}
\end{figure*}

\section{Architecture Trade Study}\label{sec:trade}
\begin{figure*}[!t]
\centering
\subfloat[Hartley model.]{\includegraphics[width=0.30\textwidth]{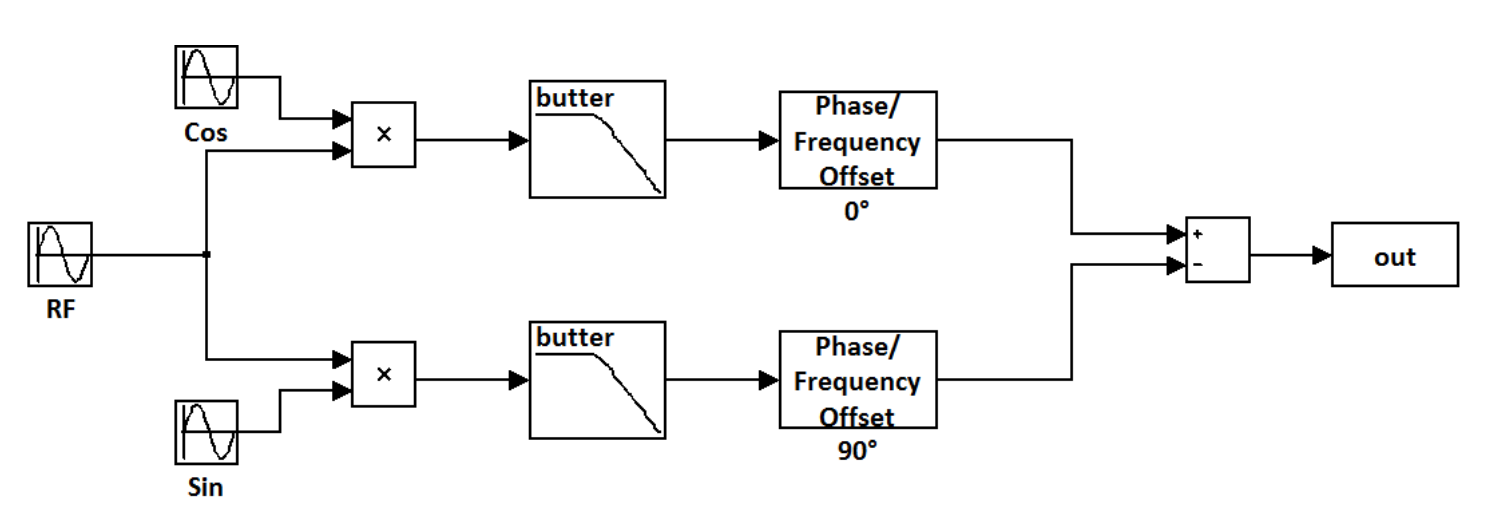}}\hfill
\subfloat[Weaver model.]{\includegraphics[width=0.31\textwidth]{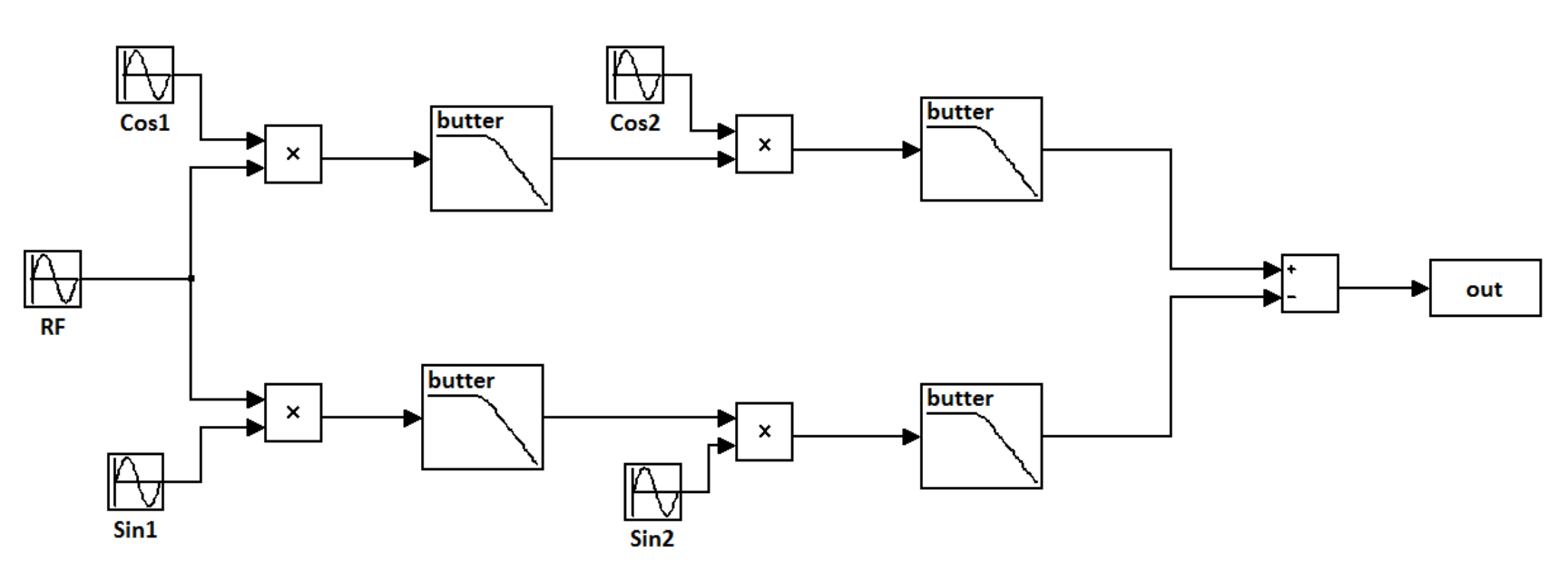}}\hfill
\subfloat[Double-conversion double-quadrature model.]{\includegraphics[width=0.32\textwidth]{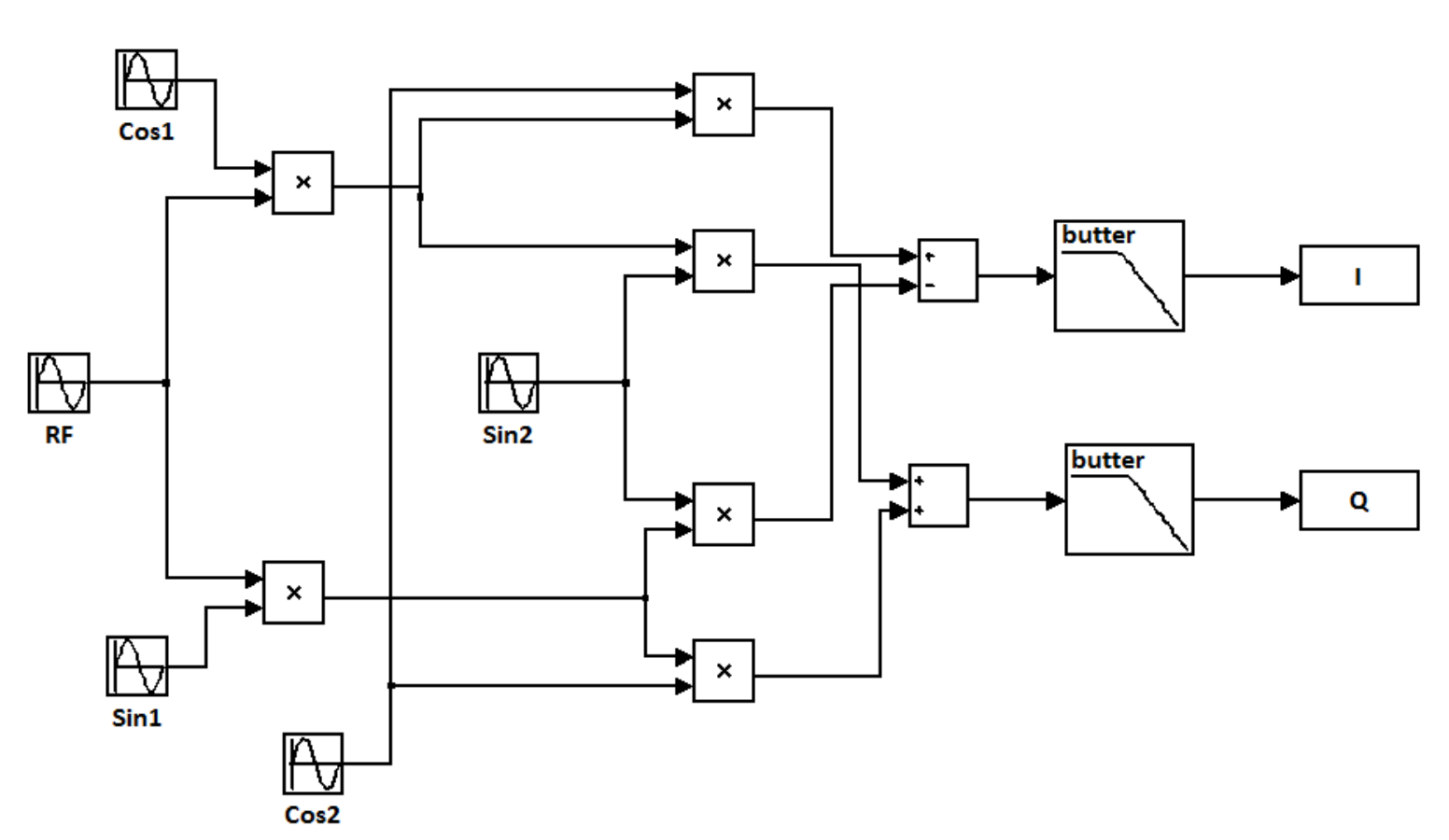}}
\caption{Behavioral Simulink models used to compare image rejection.}
\label{fig:simulink-models}
\end{figure*}

The candidate image-reject receivers were modeled in Simulink. Fourth-order Butterworth low-pass filters were used in each branch, and all discrete Fourier transforms were evaluated with the same simulation duration and sampling interval so that the numerical rejection values were comparable. The purpose of this stage was not to predict transistor-level performance, but to compare topology-dependent sensitivity under controlled I/Q perturbations; this is the same mismatch mechanism that limits practical image-reject mixers and motivates calibration in integrated receivers~\cite{behbahani2001,elahi2006}. The Simulink implementations are shown in Fig.~\ref{fig:simulink-models}.

\subsection{Sinusoidal local oscillators}
\begin{figure*}[!t]
\centering
\includegraphics[width=0.72\textwidth]{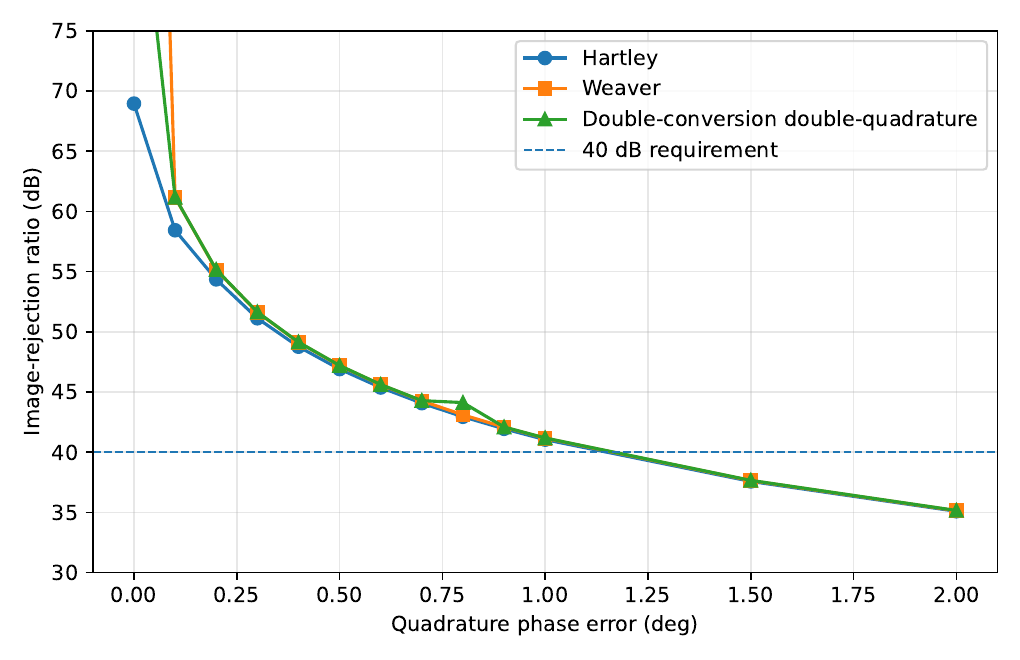}
\caption{Behavioral image rejection versus quadrature phase error with sinusoidal local oscillators. Values above \SI{75}{dB} are omitted from the plotting range because they represent ideal numerical cancellation.}
\label{fig:sinusoidal-phase}
\end{figure*}

\begin{figure*}[!t]
\centering
\includegraphics[width=0.72\textwidth]{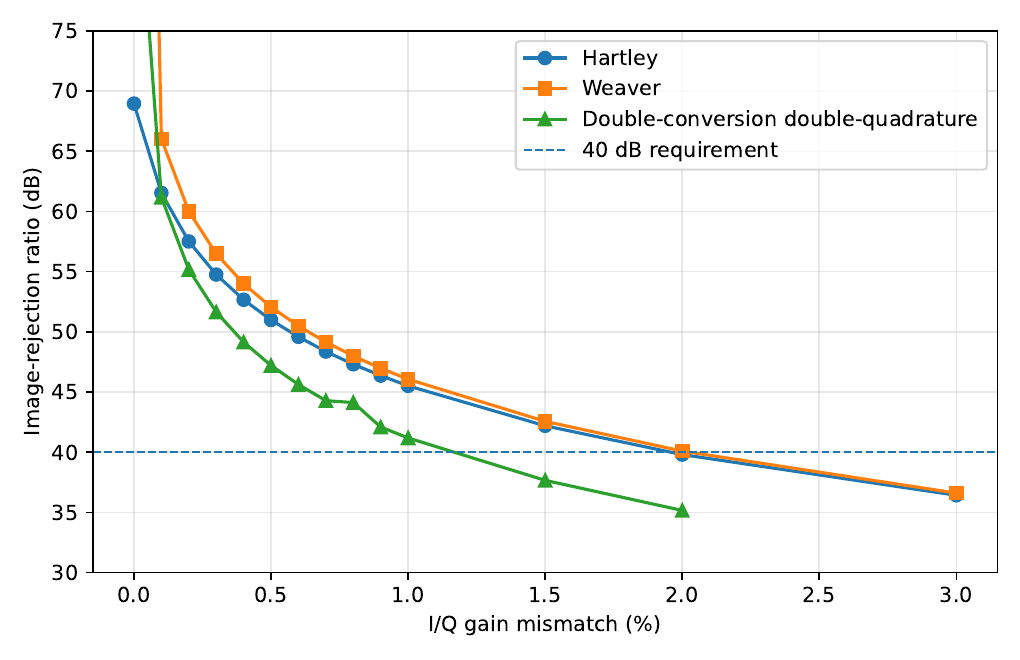}
\caption{Behavioral image rejection versus I/Q gain mismatch with sinusoidal local oscillators.}
\label{fig:sinusoidal-gain}
\end{figure*}

Figures~\ref{fig:sinusoidal-phase} and \ref{fig:sinusoidal-gain} summarize the mismatch sweeps. All three architectures show nearly identical sensitivity to phase error: image rejection remains slightly above \SI{40}{dB} at \SI{1}{\degree} and falls below the requirement for larger errors. The ideal nominal values of \SI{68.95}{dB}, \SI{170.49}{dB}, and \SI{92.23}{dB} for Hartley, Weaver, and double-quadrature structures, respectively, primarily reflect numerical cancellation limits and should not be interpreted as realizable hardware performance.

The Weaver architecture is approximately \SI{2}{dB} better than Hartley under gain mismatch and remains above \SI{40}{dB} at a \SI{2}{\percent} I/Q amplitude mismatch. The double-conversion double-quadrature results are not better and in the original table essentially track the phase-error curve, suggesting that the additional complexity is not justified by the behavioral results.

\subsection{Square-wave quadrature drive}
A continuously tuned sinusoidal VCO would require accurate quadrature generation over the full tuning range. Although sub-degree wideband quadrature generators are possible, they require dedicated analog circuitry and careful matching~\cite{kalcher2020}. Quadrature square waves are simpler to generate using digital division or timing logic and are natural drive signals for switching mixers. The resulting simplification transfers difficulty from phase generation to harmonic response: a 50\% duty-cycle square wave has the Fourier expansion
\begin{equation}
 q(t)=\frac{4}{\pi}\left[\cos(\omega t)+\frac{1}{3}\cos(3\omega t)+\frac{1}{5}\cos(5\omega t)+\cdots\right],
 \label{eq:square}
\end{equation}
so odd-harmonic mixing products must be filtered. This is a standard issue in switching-mixer receivers: modern harmonic-rejection architectures use multiphase or pulse-width-modulated LO waveforms to cancel selected harmonic responses~\cite{kang2018}. Here no such harmonic-rejection mixer is introduced; instead, the question is whether the Weaver architecture itself remains sufficiently tolerant when driven by the simpler rectangular quadrature LO.

The square-wave Hartley receiver suffers a marked reduction in nominal rejection and mismatch tolerance, whereas the Weaver receiver remains essentially unchanged. The Weaver result is physically plausible because both quadrature branches process the same harmonic family and the second conversion/filtering stage suppresses most unwanted terms. At the expected phase error below approximately \SI{0.5}{\degree}, the square-wave Weaver receiver provides about \SI{47}{dB} image rejection; it also remains above \SI{40}{dB} through a \SI{2}{\percent} gain mismatch.

\begin{figure*}[!t]
\centering
\subfloat[Phase mismatch.]{\includegraphics[width=0.47\textwidth]{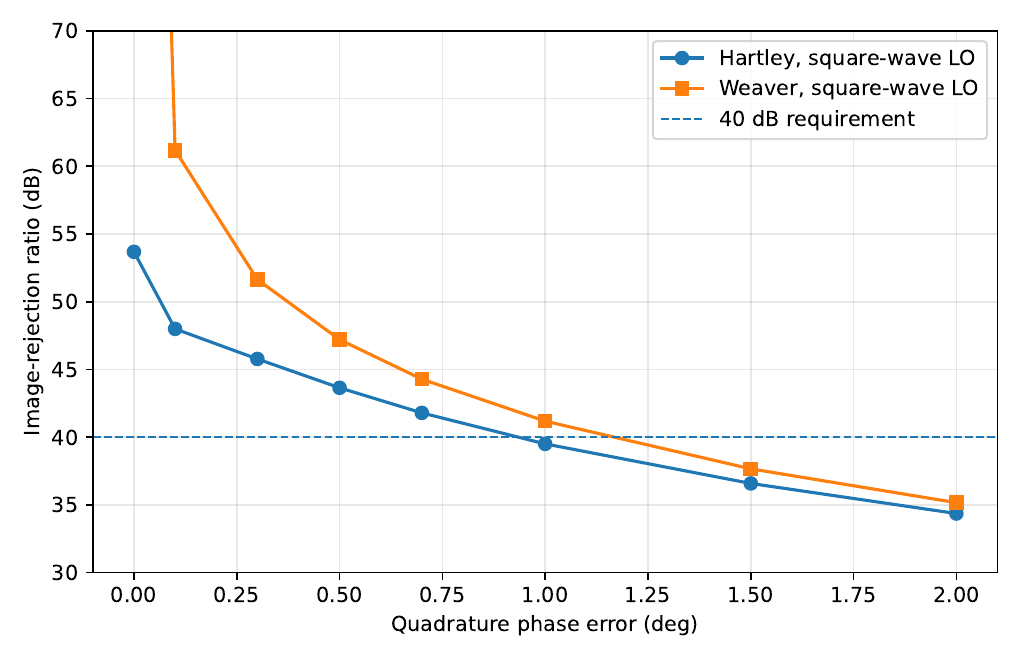}}\hfill
\subfloat[Gain mismatch.]{\includegraphics[width=0.47\textwidth]{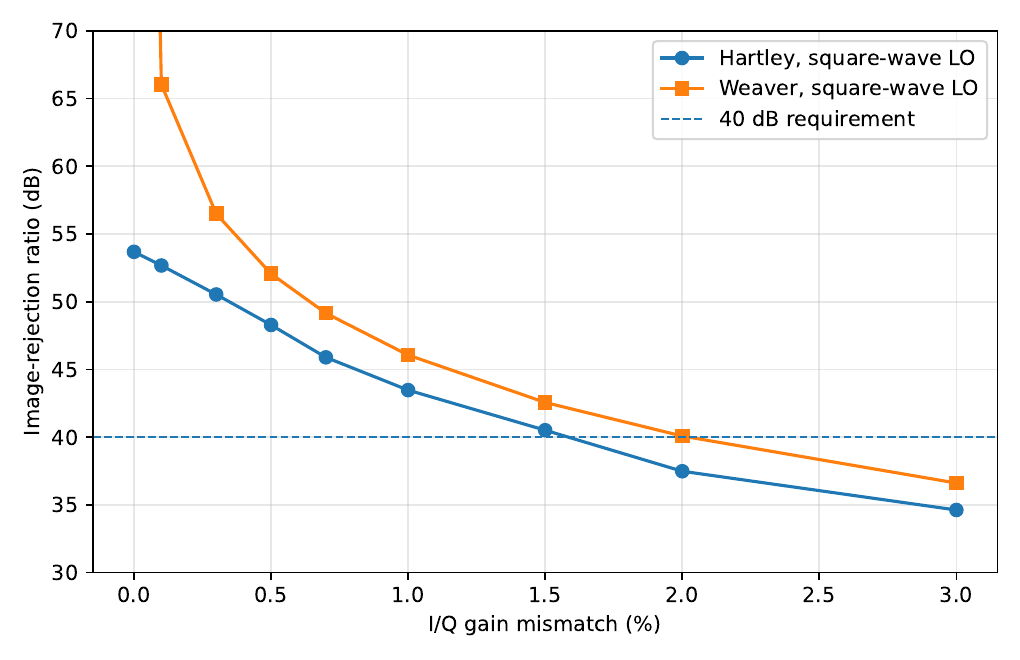}}
\caption{Behavioral comparison of Hartley and Weaver receivers using square-wave quadrature local oscillators.}
\label{fig:square-results}
\end{figure*}

\section{Frequency-Agile Weaver Channelizer}\label{sec:weaver}
\subsection{Halving the variable-oscillator tuning span}
The second key design problem is the \SI{50}{MHz} tuning span of a single oscillator. The proposed solution exploits the sideband-selection property inherent to quadrature image-reject conversion~\cite{weaver1956,behbahani2001}: the second LO is fixed, while the output combiner selects one of two RF frequencies that are symmetric about the first-LO frequency. The input band is divided into lower and upper \SI{25}{MHz} halves. For any first-LO setting, a frequency in one half of the band is the image partner of a frequency in the other half. Because the I branch contains the desired and image terms with the same sign while the Q branch contains them with opposite signs, output subtraction selects one member of the pair and output addition selects the other.

The first conversion maps the selected \SI{3}{MHz} slice to an intermediate band from \SI{23}{MHz} to \SI{26}{MHz}. A fixed sinusoidal second LO at \SI{26}{MHz} then translates this band to dc--\SI{3}{MHz}. Only the first LO must be tunable, and its required span is reduced to \SI{25}{MHz}. The second LO is fixed-frequency and can therefore be sinusoidal, avoiding its own square-wave harmonic products.

\begin{figure*}[!t]
\centering
\includegraphics[width=0.72\textwidth]{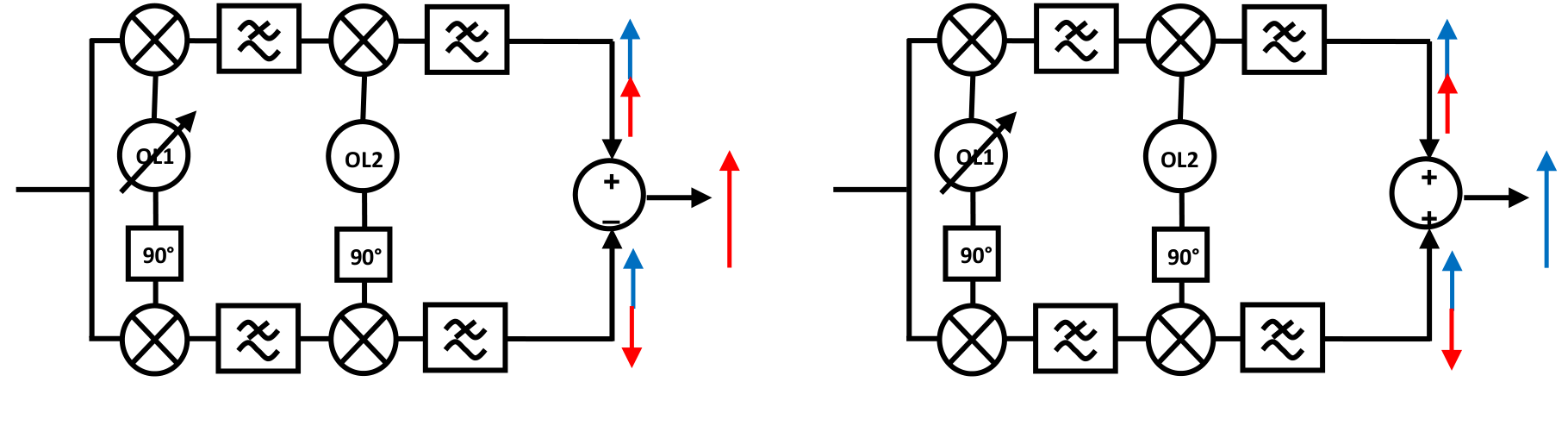}
\caption{Band-selection principle. The two half-bands form image pairs. Subtraction selects one sideband and summation selects the other, allowing the variable oscillator to cover only half of the original input span.}
\label{fig:final-architecture}
\end{figure*}

\subsection{Frequency-plan implementation considerations}
The band-partitioning argument establishes frequency coverage, but a hardware implementation must additionally define the exact first-LO tuning law, the mapping from requested RF channel to combiner polarity, switching transients, and anti-aliasing constraints. These control-layer details do not affect the sideband-selection principle demonstrated here, but they determine channel-to-channel settling time and should be included in any subsequent integrated implementation.

\section{Circuit-Level Realization}\label{sec:circuit}
The transistor-level receiver was implemented in Cadence Virtuoso and simulated with Spectre. The circuit partition follows a conventional integrated-receiver philosophy in which multiplication and low-pass selection are implemented with CMOS transconductance/mixer stages and active filters rather than external high-$Q$ networks~\cite{behbahani2001,russell2020}. The available design record does not include the CMOS process node, supply voltage, device dimensions, bias currents, total power, or silicon area. Consequently, the circuit results below are reported as schematic-level functional validation and are not used for process-normalized power, area, or noise comparisons with measured receiver ICs.

\subsection{Operational amplifier}
The fourth-order active filters require an operational amplifier with sufficient dc gain, bandwidth, and phase margin. The implemented amplifier comprises an enable/bias stage, a differential high-gain stage with active load, and a common-source output stage. Miller compensation is applied between the second and third stages to create a dominant pole~\cite{sedra2004,razavi2006}.

\begin{figure*}[!t]
\centering
\subfloat[Feedback-system model.]{\includegraphics[width=0.25\textwidth]{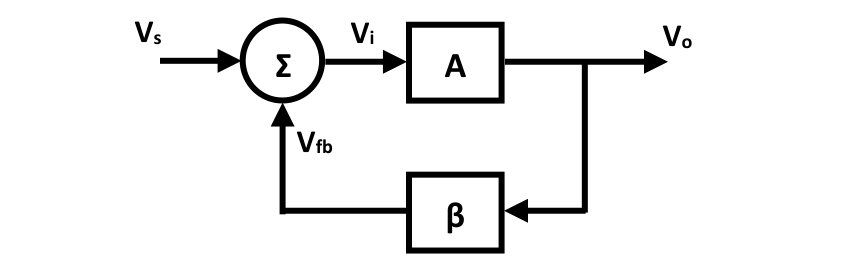}}\hfill
\subfloat[Enable and bias stage.]{\includegraphics[width=0.31\textwidth]{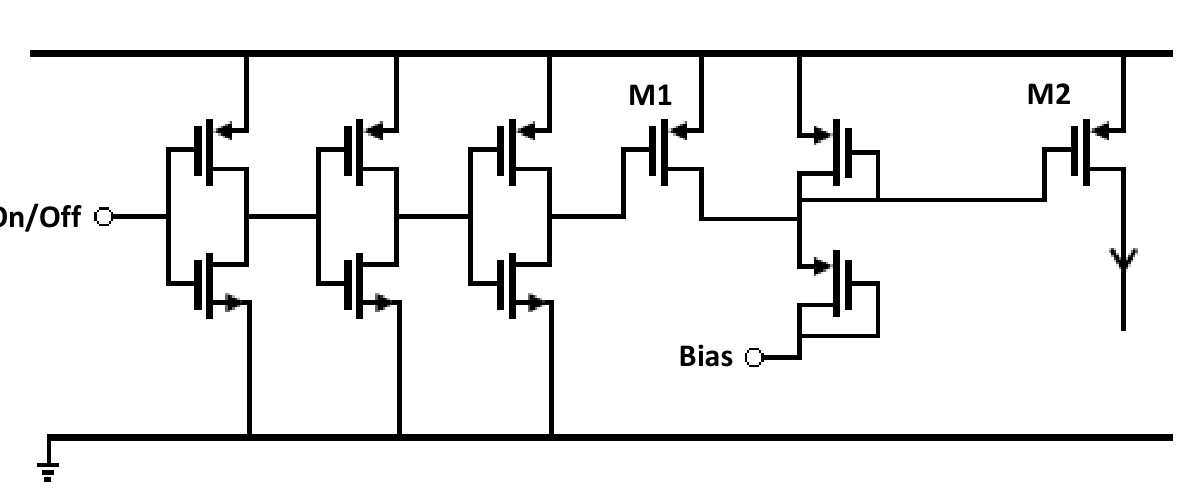}}\hfill
\subfloat[Differential high-gain stage.]{\includegraphics[width=0.22\textwidth]{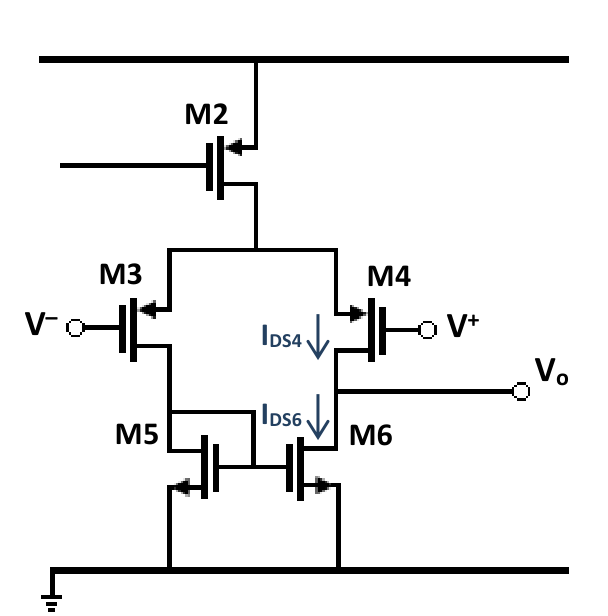}}\\[6pt]
\subfloat[MOS small-signal model.]{\includegraphics[width=0.19\textwidth]{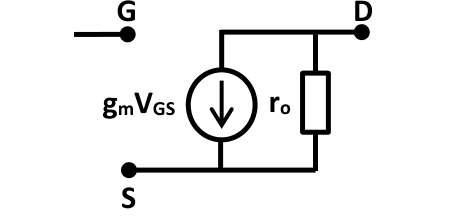}}\hfill
\subfloat[Common-source output stage.]{\includegraphics[width=0.16\textwidth]{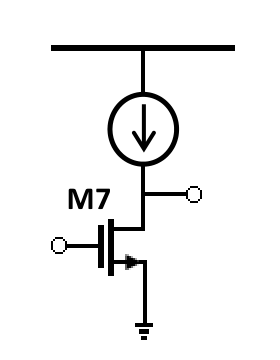}}\hfill
\subfloat[Equivalent small-signal output stage.]{\includegraphics[width=0.25\textwidth]{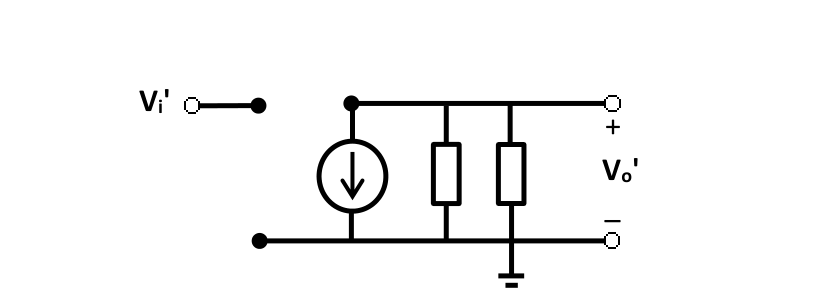}}
\caption{Operational-amplifier building blocks and small-signal models.}
\label{fig:opamp-building-blocks}
\end{figure*}

For the feedback system in Fig.~\ref{fig:opamp-building-blocks}(a), the closed-loop transfer function is
\begin{equation}
 H(s)=\frac{A(s)}{1+A(s)\beta(s)}.
\end{equation}
Stability requires the loop-gain magnitude to fall below unity before its phase reaches \SI{-180}{\degree}. A phase margin close to \SI{60}{\degree} was used as the design target.

The small-signal gain of the differential second stage is approximated by
\begin{equation}
 A_2\simeq -g_{m3}\left(r_{o4}\parallel r_{o6}\right),
 \label{eq:a2}
\end{equation}
and the common-source third-stage gain is
\begin{equation}
 A_3\simeq -g_{m7}\left(r_{o7}\parallel r_{o8}\right).
 \label{eq:a3}
\end{equation}
The complete open-loop gain is the product of the stage gains, modified by loading and compensation.

\begin{figure*}[!t]
\centering
\subfloat[Complete operational-amplifier schematic.]{\includegraphics[width=0.47\textwidth]{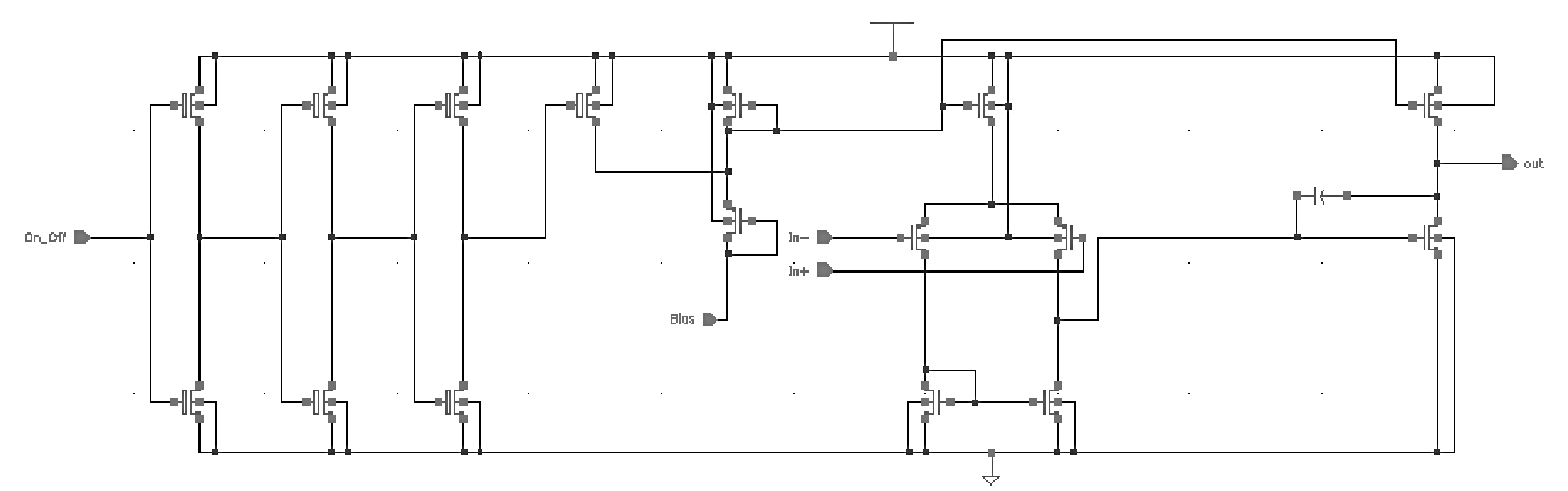}}\hfill
\subfloat[Open-loop phase and gain response.]{\includegraphics[width=0.47\textwidth]{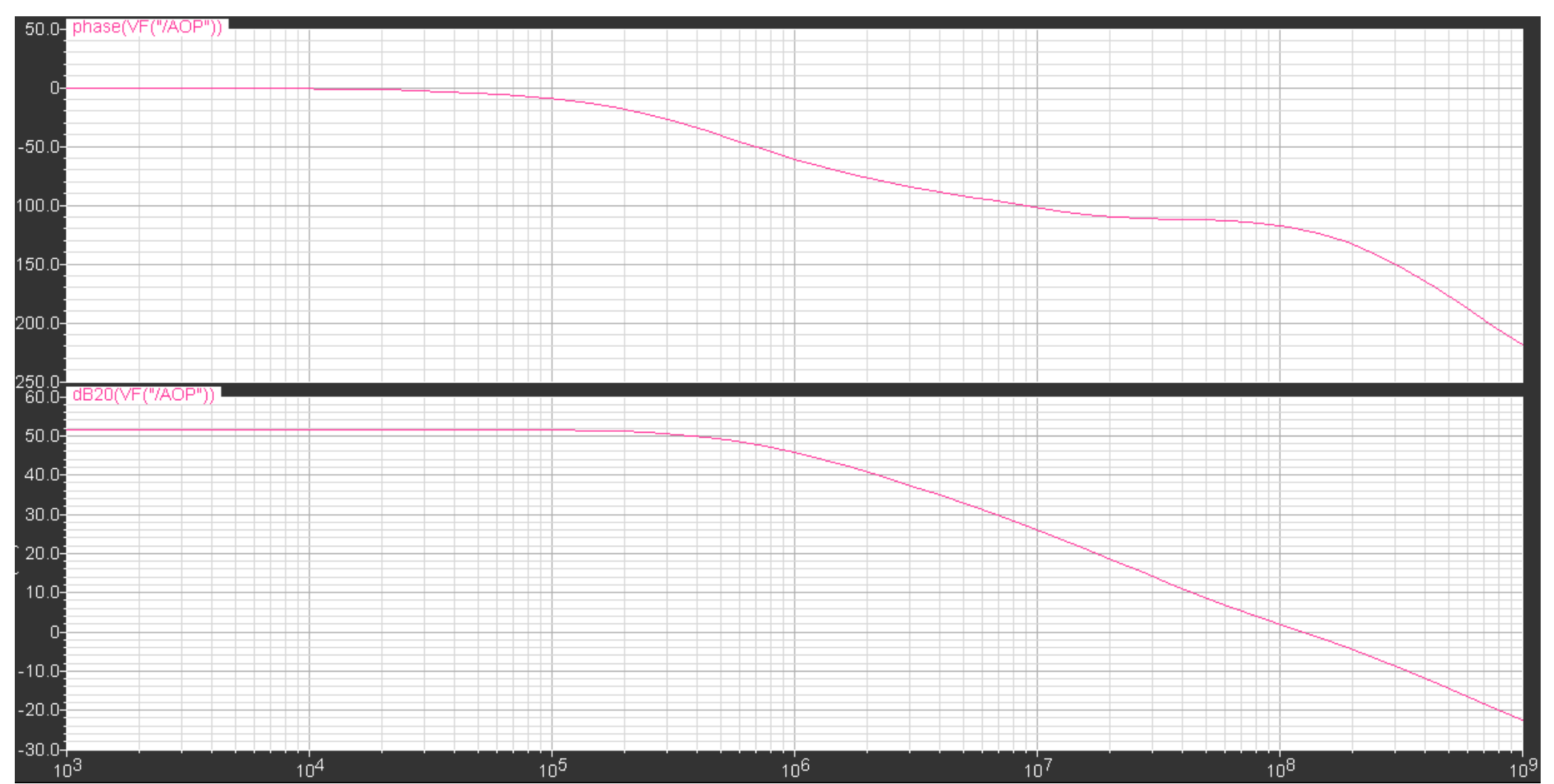}}
\caption{Complete operational amplifier and its nominal simulated frequency response.}
\label{fig:opamp-complete}
\end{figure*}

The complete schematic and simulated Bode response are shown in Fig.~\ref{fig:opamp-complete}. The low-frequency gain is approximately \SI{52}{dB}; the response reaches \SI{0}{dB} near \SI{200}{MHz}; and the simulated phase margin is \SI{62}{\degree}. The simulated response is approximately flat through \SI{130}{kHz}. These values are adequate for the intended low-frequency active filters, although a complete stability study would also require process, voltage, temperature, capacitive-load, and common-mode sweeps.

\subsection{Mixer}
The mixer is based on a Gilbert-cell topology~\cite{lee2004,razavi2006}, a standard choice when transconductance multiplication and differential LO switching are required, with two modifications. First, the input transconductor uses an input-dependent bias current to increase the usable input range. Second, a current-mirror output stage converts the differential mixer current to a single-ended signal so that conventional single-ended active filters can be used.

If the positive and negative RF inputs have equal amplitude and opposite phase, the variable tail-current stage produces approximately
\begin{equation}
 I_0=KC^2\left(v_{\mathrm{RF}+}^2+v_{\mathrm{RF}-}^2\right)+2KV_t^2,
 \label{eq:tail}
\end{equation}
where $K$ is the MOS square-law factor, $C$ is the input scaling factor, and $V_t$ is the threshold voltage. The linear terms cancel because $v_{\mathrm{RF}+}+v_{\mathrm{RF}-}=0$.

For a differential pair with tail current $I_{SS}$ and small differential input $v_{id}$, the output-current difference is approximated by
\begin{equation}
 i_o\simeq \sqrt{K I_{SS}}\,v_{id}.
 \label{eq:diffpair}
\end{equation}
Applying this relation to the RF transconductor and LO switching pairs yields
\begin{equation}
 i_{\mathrm{IF}}\simeq K\left(v_{\mathrm{RF}+}-v_{\mathrm{RF}-}\right)
 \left(v_{\mathrm{LO}+}-v_{\mathrm{LO}-}\right),
 \label{eq:gilbert}
\end{equation}
which is proportional to the RF--LO product. The single-ended mirror stage ideally preserves this difference current and can also provide current gain.

\begin{figure*}[!t]
\centering
\subfloat[Input-dependent bias and current-mirror stage.]{\includegraphics[width=0.31\textwidth]{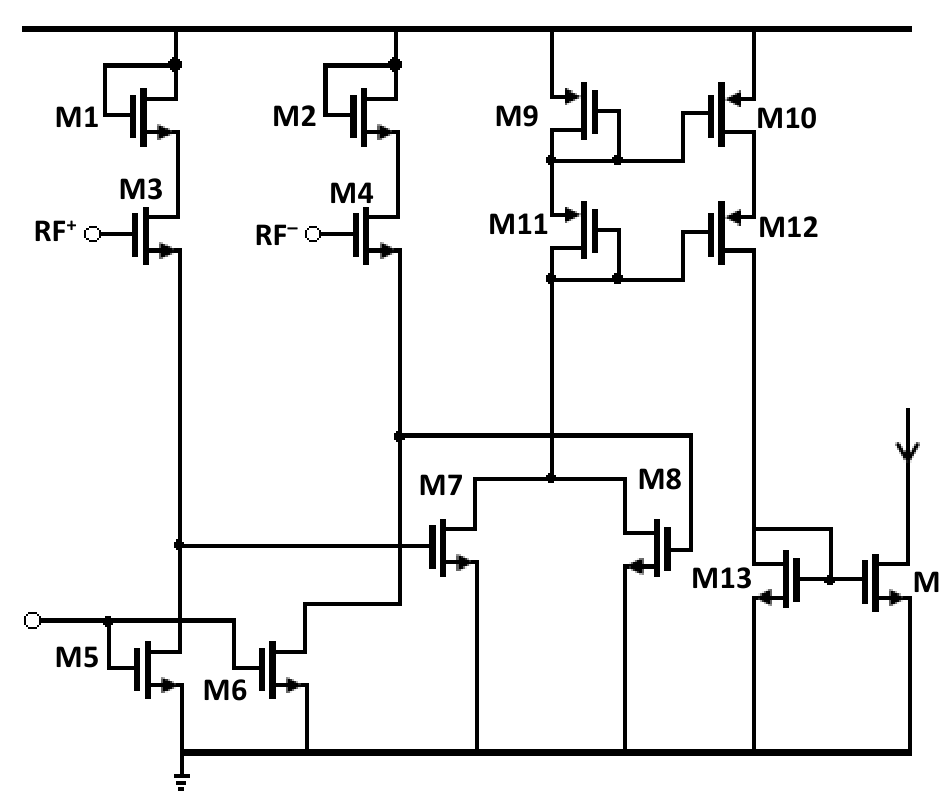}}\hfill
\subfloat[Gilbert-cell mixing core.]{\includegraphics[width=0.24\textwidth]{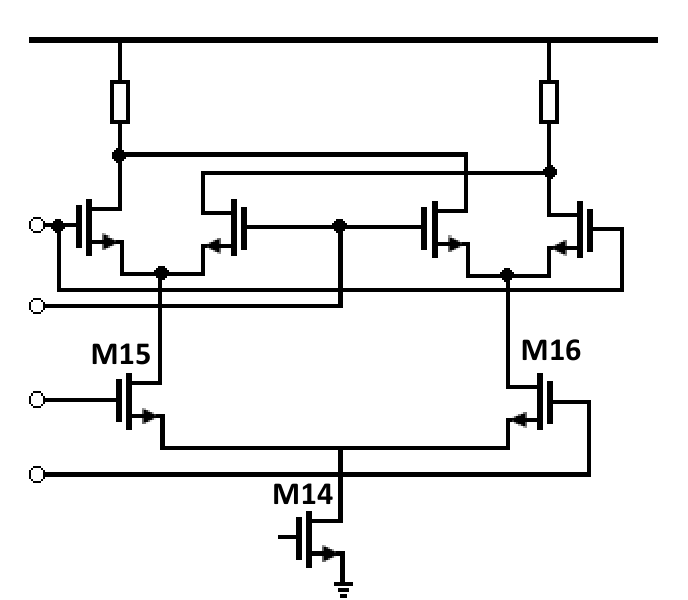}}\hfill
\subfloat[Differential-pair model.]{\includegraphics[width=0.16\textwidth]{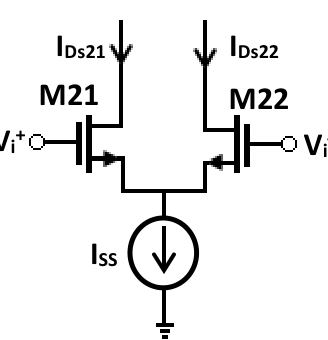}}\\[6pt]
\subfloat[Single-ended output conversion stage.]{\includegraphics[width=0.29\textwidth]{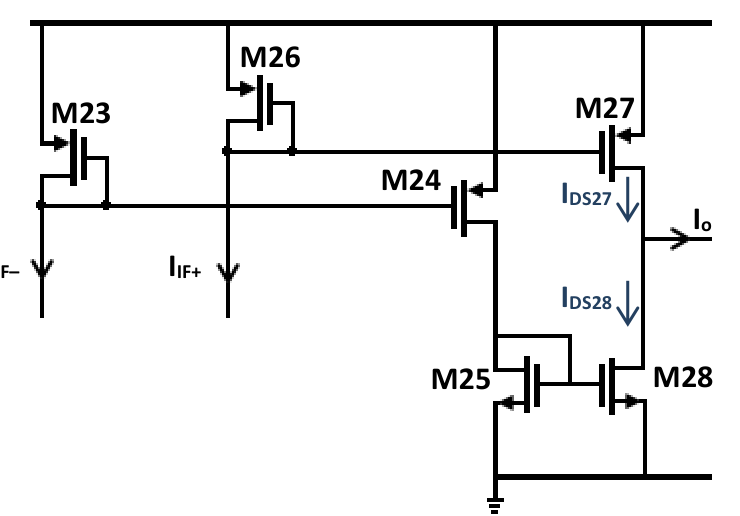}}
\caption{Mixer subcircuits used in the transistor-level receiver.}
\label{fig:mixer-stages}
\end{figure*}

\begin{figure*}[!t]
\centering
\subfloat[Complete mixer schematic.]{\includegraphics[width=0.47\textwidth]{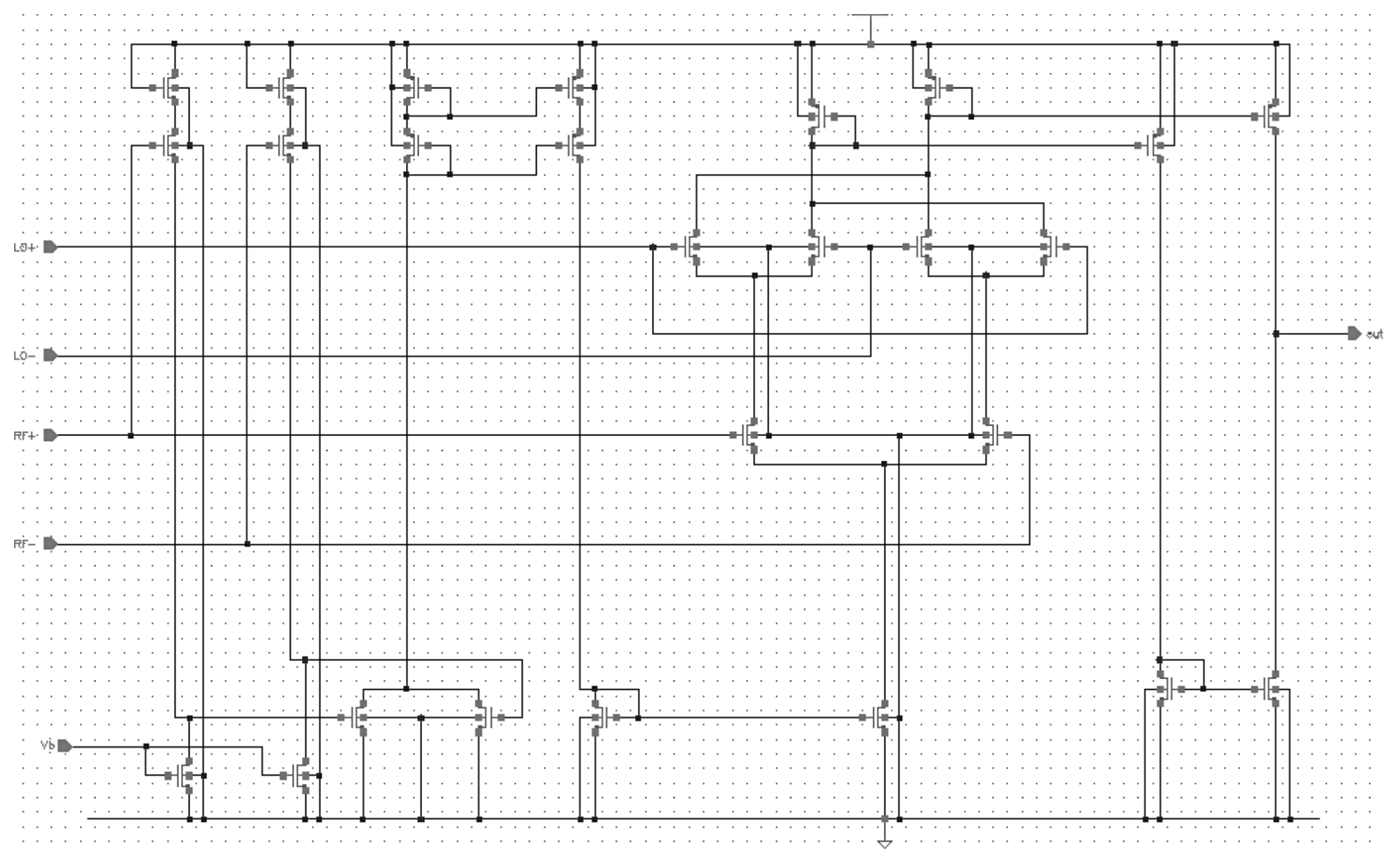}}\hfill
\subfloat[Differential output spectrum of the mixing core.]{\includegraphics[width=0.47\textwidth]{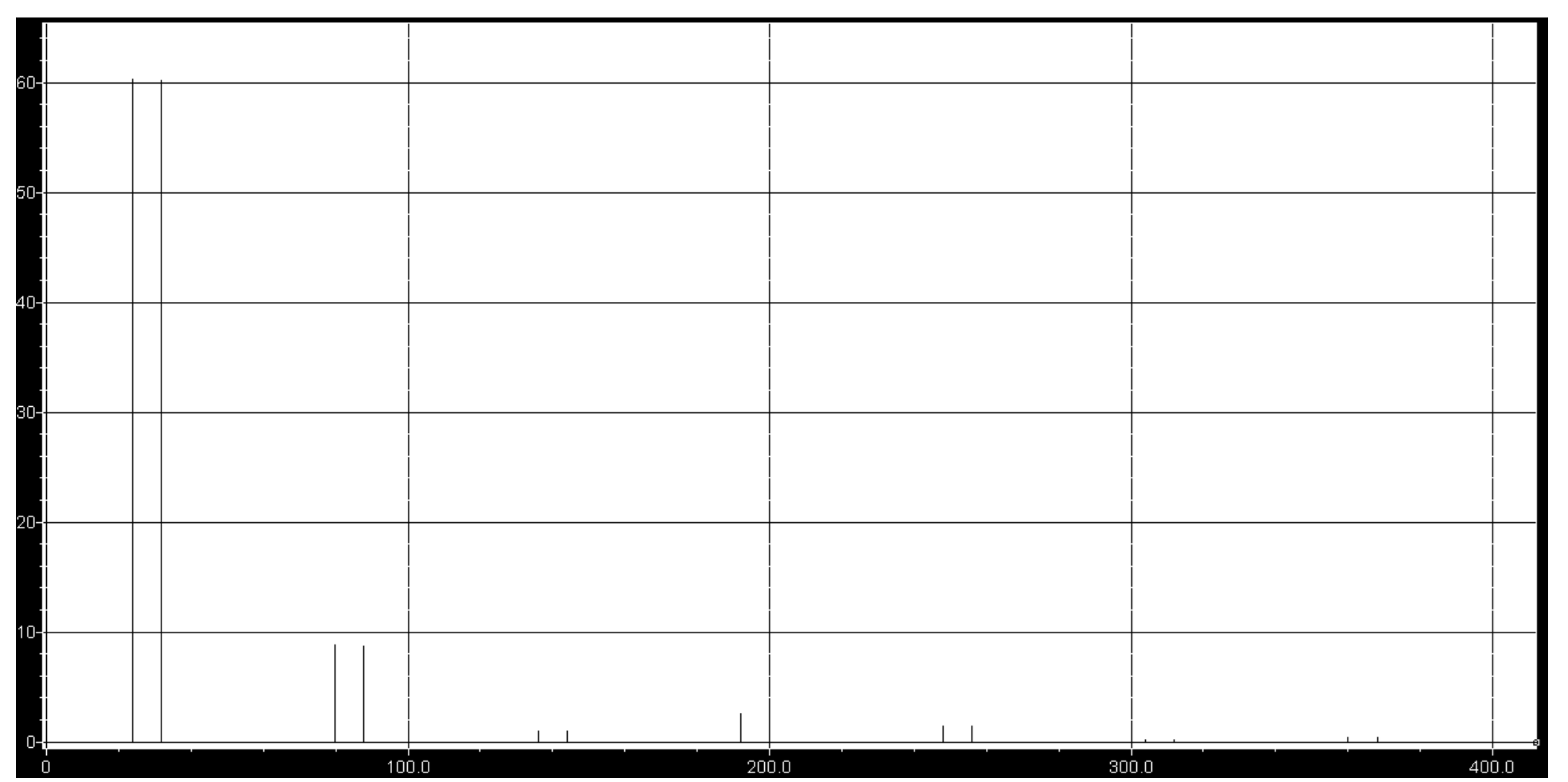}}\\[6pt]
\subfloat[Single-ended mixer output spectrum.]{\includegraphics[width=0.58\textwidth]{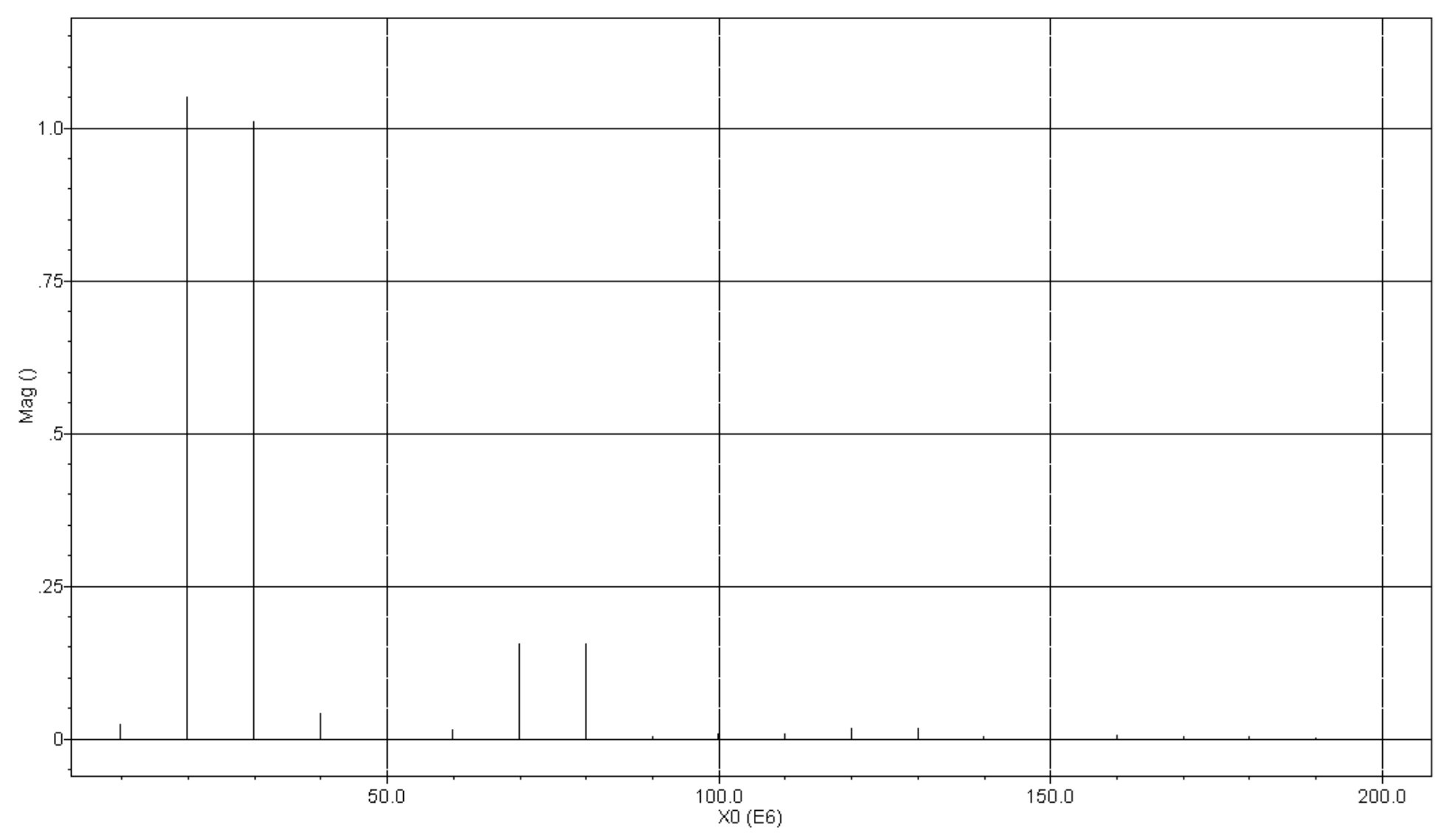}}
\caption{Complete mixer and representative spectra under square-wave local-oscillator drive.}
\label{fig:mixer-complete}
\end{figure*}

The complete mixer schematic and simulated spectra are shown in Fig.~\ref{fig:mixer-complete}. With square-wave LO drive, odd-harmonic mixing products appear at combinations of $(2m+1)\omega_{\mathrm{LO}}\pm\omega_{\mathrm{RF}}$. This behavior follows directly from the Fourier content of a switching LO and is the same mechanism addressed by harmonic-rejection mixer techniques~\cite{kang2018}. The second-stage differential spectrum shows these components explicitly. After single-ended conversion, the desired component is dominant, with weaker parasitic lines. The simulated conversion gain is approximately \SI{7}{dB}; the ratio between the desired component at $\omega_{\mathrm{LO}}-\omega_{\mathrm{RF}}$ and the component at $3(\omega_{\mathrm{LO}}-\omega_{\mathrm{RF}})$ is approximately \SI{18}{dB} before the complete receiver filtering.

\subsection{Fourth-order active filter}
Each low-pass filter is implemented as a cascade of two active second-order sections. Active filtering is particularly natural at these low IFs because it avoids the inductors and high-$Q$ resonators that complicate monolithic image filtering~\cite{behbahani2001}. Component values were chosen to keep resistances in the tens-of-kilohms range and capacitances between hundreds of femtofarads and tens of picofarads, making monolithic integration plausible. The nonideal operational amplifier introduces a resonance relative to the ideal response. The approximately \SI{5}{MHz}-wide enhanced-gain region is aligned with the translated \SI{3}{MHz} channel to increase the selected-band amplitude relative to the rest of the spectrum. This intentional peaking improves selectivity in the nominal response but can worsen transient behavior and PVT sensitivity, so its robustness requires further validation.

\begin{figure*}[!t]
\centering
\subfloat[Fourth-order active-filter topology.]{\includegraphics[width=0.42\textwidth]{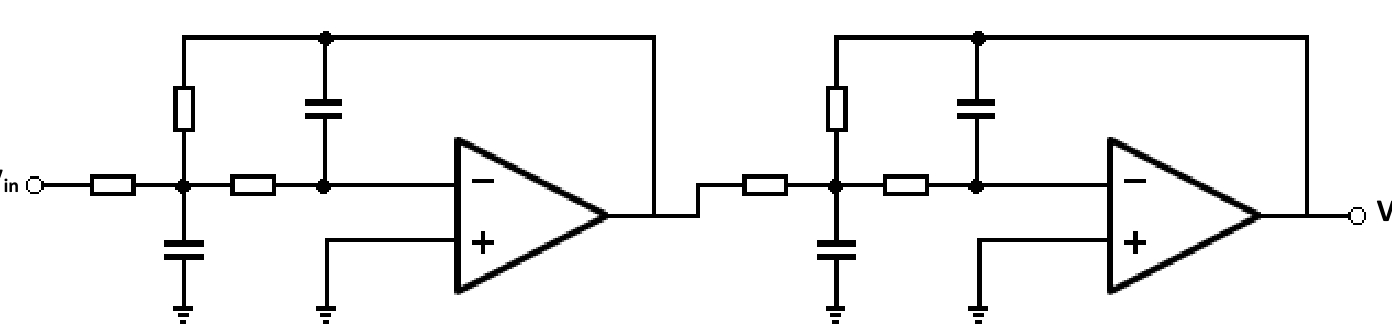}}\hfill
\subfloat[Real and ideal filter responses.]{\includegraphics[width=0.47\textwidth]{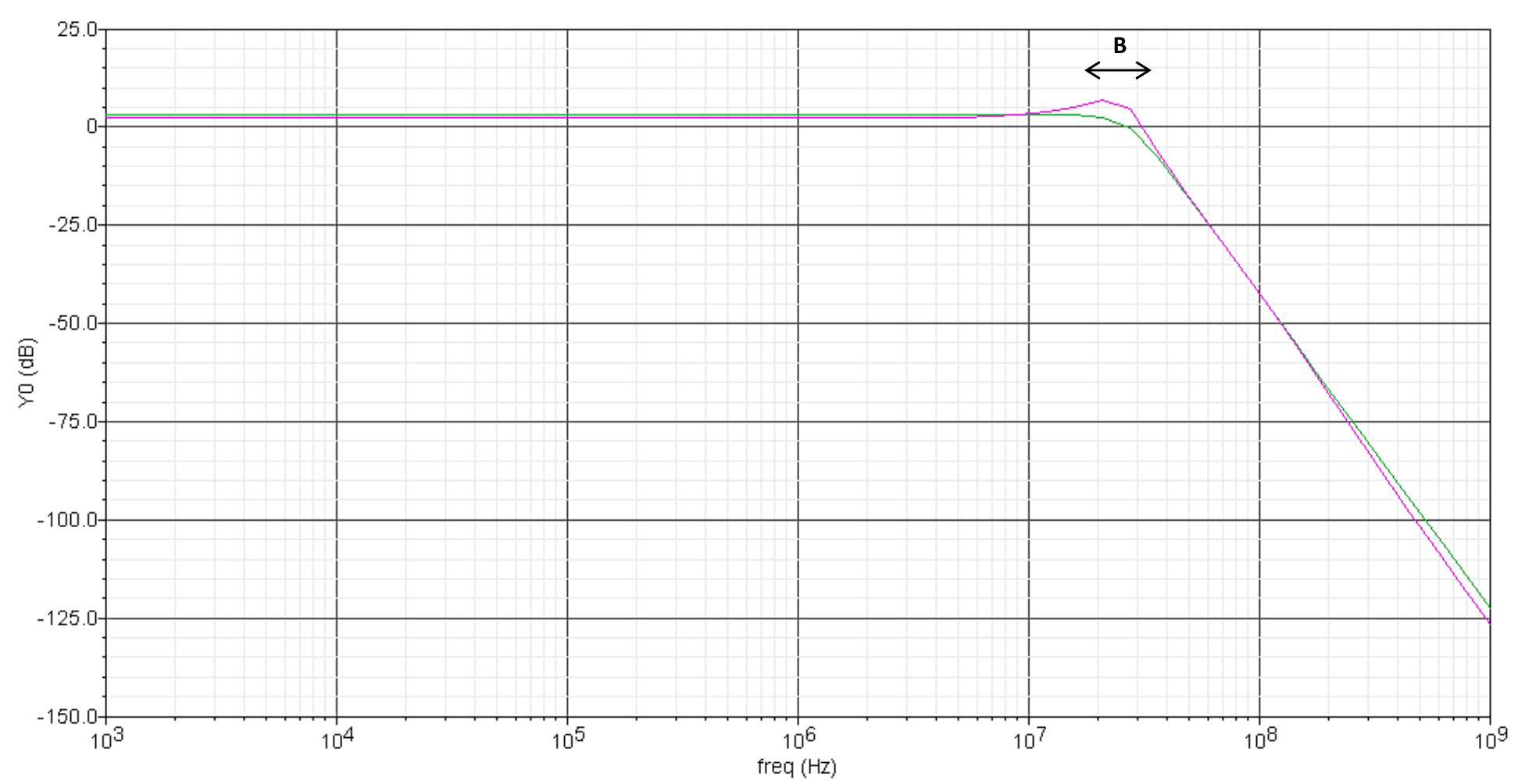}}\\[6pt]
\subfloat[End-to-end receiver output spectrum for a \SI{50}{MHz} input translated to \SI{1}{MHz}.]{\includegraphics[width=0.62\textwidth]{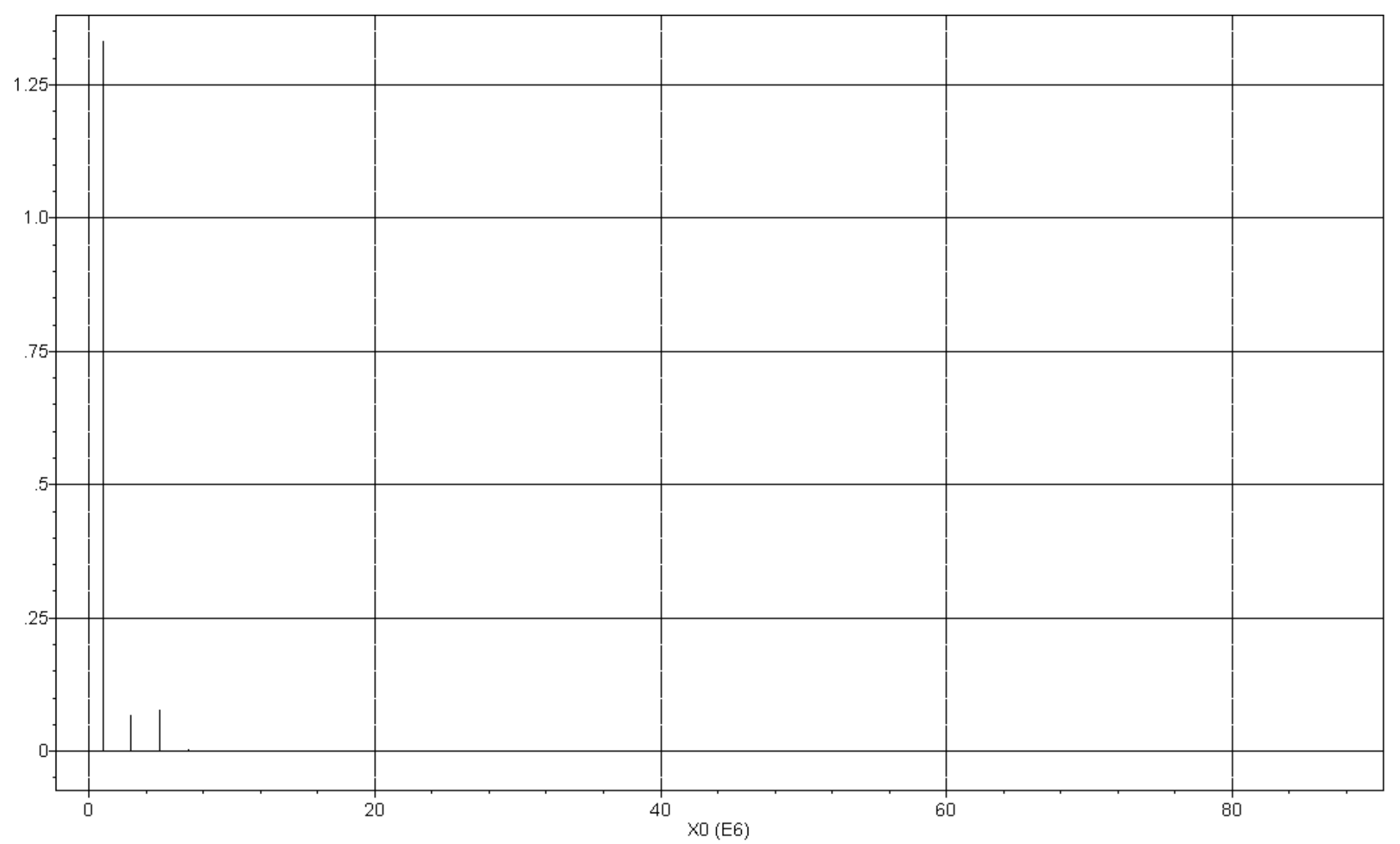}}
\caption{Filter implementation and end-to-end receiver output spectrum.}
\label{fig:filter-final-spectrum}
\end{figure*}

\section{End-to-End Receiver Performance}\label{sec:performance}
\subsection{Output spectrum and harmonics}
For a \SI{50}{MHz} input, the receiver translates the desired signal to approximately \SI{1}{MHz}. The output spectrum in Fig.~\ref{fig:filter-final-spectrum}(c) contains harmonics near \SI{3}{MHz} and \SI{5}{MHz}. The desired-to-\SI{3}{MHz} ratio is approximately \SI{27}{dB}, and the desired-to-\SI{5}{MHz} ratio is approximately \SI{25.5}{dB}. Increasing the filter order would improve harmonic suppression, but at the cost of additional area, power, sensitivity to component spread, and stability burden. Alternatively, the LO waveform itself could be shaped using harmonic-rejection techniques, as in PWM-based or multiphase switching receivers~\cite{kang2018}. The low operating frequencies make the filtering problem particularly demanding because the odd-harmonic products remain close to the selected band.

\subsection{Quadrature phase error}
The complete transistor-level receiver remains above the \SI{40}{dB} requirement over the entire simulated range through \SI{1}{\degree} phase error. The steep degradation is consistent with the standard I/Q-mismatch dependence in (\ref{eq:irr-mismatch}) and with the broader image-reject receiver literature, in which sub-degree phase control or calibration is typically required once the desired IRR approaches several tens of decibels~\cite{behbahani2001,kalcher2020}. The nominal value is \SI{71.43}{dB}; the rejection is \SI{49.97}{dB} at \SI{0.5}{\degree} and \SI{42.89}{dB} at \SI{1}{\degree}. The design estimate for the maximum square-wave quadrature error was below \SI{0.5}{\degree}; at that error the simulated receiver retains roughly a \SI{10}{dB} margin over the requirement.

\begin{figure*}[!t]
\centering
\includegraphics[width=0.68\textwidth]{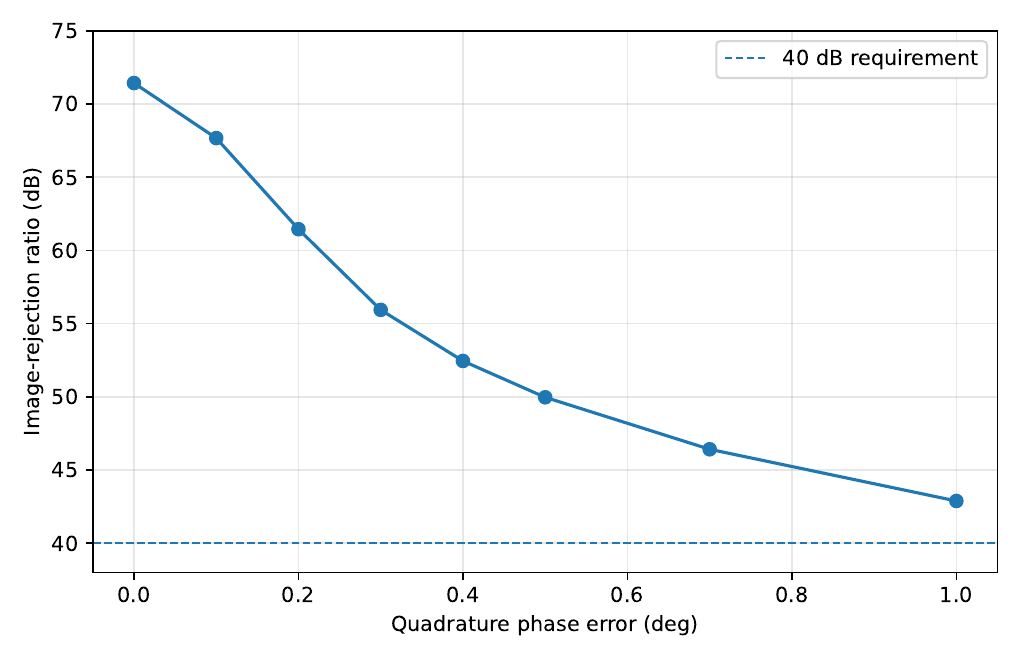}
\caption{End-to-end receiver image rejection versus quadrature phase error.}
\label{fig:final-phase}
\end{figure*}

\subsection{Process corners}
Three representative process-corner combinations were evaluated, with resistor and capacitor models held at their worst-speed corners. The image rejection is \SI{65.04}{dB} with typical CMOS devices, \SI{58.00}{dB} with worst-power CMOS devices, and \SI{43.73}{dB} with all device classes at their worst-speed corners. Thus, the simulated corner set remains above the requirement, although only narrowly in the all-worst-speed case. These deterministic corners do not replace Monte Carlo mismatch analysis, which is especially important for I/Q amplitude and phase balance~\cite{weste2005,behbahani2001}. Published high-IRR receivers commonly rely on either careful passive matching or explicit calibration precisely because systematic PVT corners do not capture local random mismatch~\cite{der2003,ma2023}.

\begin{figure*}[!t]
\centering
\includegraphics[width=0.72\textwidth]{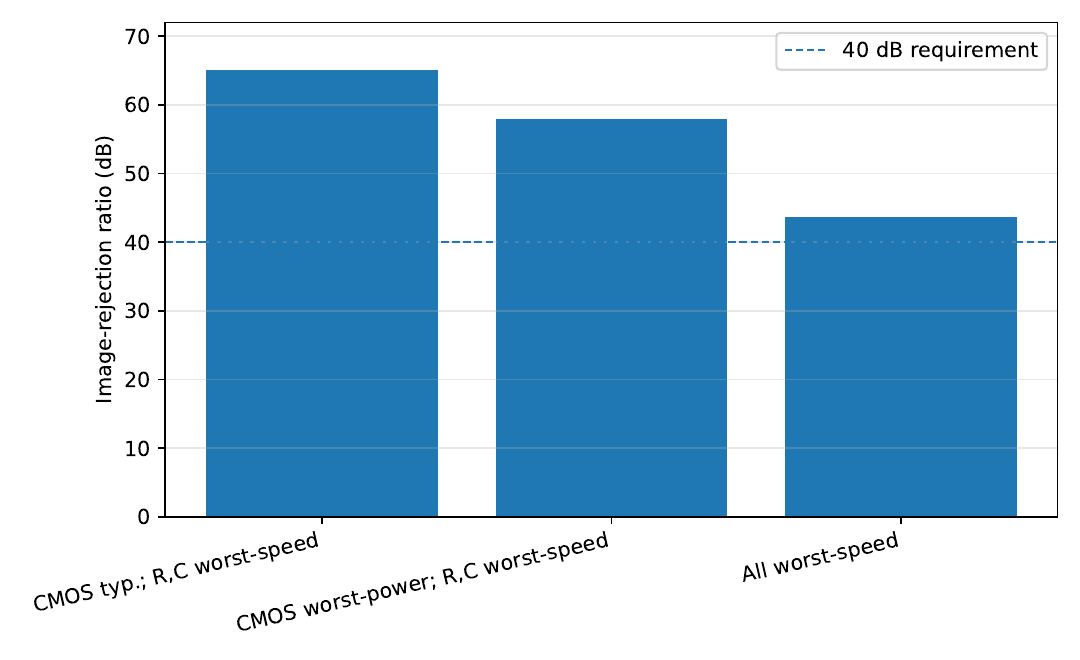}
\caption{End-to-end receiver image rejection for the three evaluated process-corner combinations.}
\label{fig:corners}
\end{figure*}

\subsection{Temperature dependence}
The simulated image rejection peaks near \SI{0}{\celsius} and decreases rapidly above \SI{40}{\celsius}. It is \SI{45.14}{dB} at \SI{50}{\celsius} and \SI{38.43}{dB} at \SI{60}{\celsius}; linear interpolation places the \SI{40}{dB} crossing near \SI{58}{\celsius}. The high-temperature degradation is consistent with worsening I/Q gain and phase balance. Although the receiver meets the original requirement over much of the simulated range, the result does not establish operation over a standard space-qualified temperature range.

\begin{figure*}[!t]
\centering
\includegraphics[width=0.70\textwidth]{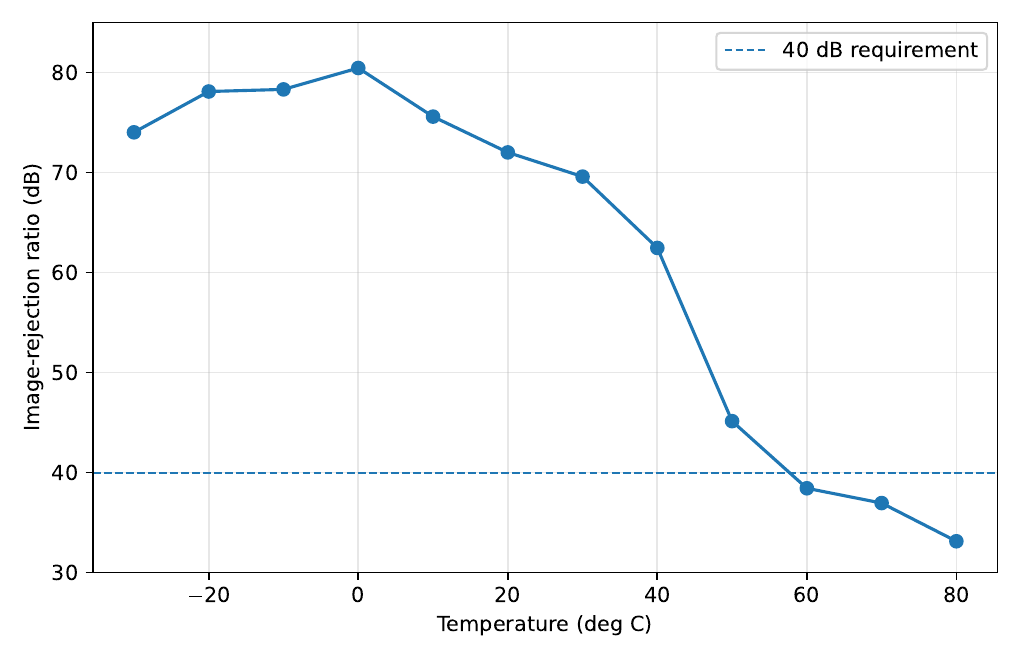}
\caption{End-to-end receiver image rejection versus temperature.}
\label{fig:temperature}
\end{figure*}

\section{Instrument-Level Discussion}\label{sec:discussion}
\subsection{What the receiver architecture establishes}
The principal instrument-level result is a frequency-selective analog channelizer that covers a \SIrange{0}{50}{MHz} input band while requiring only a \SI{25}{MHz} tuning span in the variable first LO.  The Weaver combination performs sideband selection without a high-$Q$ RF image filter, and the second conversion delivers a \SIrange{0}{3}{MHz} output suitable for subsequent digitization and spectral estimation.  In that sense, the design addresses a recurring problem in compact low-frequency radio payloads: how to trade analog selectivity, LO complexity, and digital bandwidth against one another under spacecraft resource constraints.  The frequency range is directly relevant to current low-frequency instrument concepts, including the proposed \SI{50}{kHz}--\SI{50}{MHz} RAISE payload and lunar concepts operating through roughly \SI{50}{MHz}~\cite{raise2026,farview2026}.

The nominal \SI{71.43}{dB} image-rejection value should not be interpreted as the robust performance level of the receiver.  The stressed cases are more informative: \SI{42.89}{dB} at \SI{1}{\degree} phase error and \SI{43.73}{dB} in the all-worst-speed corner leave only \SI{2.89}{dB} and \SI{3.73}{dB} of margin above the design requirement.  The temperature sweep similarly shows that the margin collapses above roughly \SI{50}{\celsius}.  The end-to-end simulations therefore identify I/Q phase and gain balance as the dominant control variable, consistent with measured low-IF and Weaver receivers in which quadrature accuracy or explicit calibration sets the achievable IRR~\cite{behbahani2001,der2003,russell2020,ma2023}.

\subsection{Relation to an astrophysical measurement chain}
Image rejection is only one element of a radio instrument's sensitivity and calibration budget.  To convert the output power spectrum into physical flux density or electric-field spectral density, a flight instrument also requires a calibrated antenna effective length, front-end transfer function, receiver noise model, and stable gain reference.  Solar Orbiter/RPW provides a representative example: in-flight antenna and receiver calibration is required to convert measured voltage spectra into astrophysical radio fluxes~\cite{vecchio2021}.  Because input matching, noise figure, absolute gain calibration, and antenna response are not available for the present design, no system-equivalent flux density or sky-noise-limited sensitivity is claimed here.

The \SI{3}{MHz} instantaneous output bandwidth also makes the architecture a swept or stepped spectrometer front end rather than an instantaneous \SI{50}{MHz} recorder.  That can reduce ADC sampling rate and downstream digital processing, but it introduces a separate science trade: the LO settling time, dwell time, and sweep cadence determine whether rapidly evolving structures in solar radio bursts are adequately sampled.  Those timing quantities were not part of the available circuit simulations and must be specified at instrument level for a particular science case.  Conversely, for observations that dwell on selected sub-bands or for slowly varying spectral backgrounds, the reduced instantaneous bandwidth can be an advantage.

Image rejection should likewise be distinguished from radio-frequency-interference rejection.  The Weaver cancellation suppresses the frequency-conversion image, but a strong interferer that lies inside the selected RF sub-band remains in band.  This distinction is particularly important for near-Earth low-frequency payloads, where modern concepts explicitly treat the RFI environment as part of the instrument design problem~\cite{raise2026}.  Preselection, linearity, gain control, and digital excision therefore remain necessary in a complete payload.

\subsection{Validation boundaries and next instrument steps}
The validation presented here is schematic-level.  Noise figure, sensitivity, dynamic range, \SI{1}{dB} compression, IIP2/IIP3, LO leakage, dc offset, and channel-switching settling time were not evaluated.  The oscillator circuits were not implemented at transistor level, and no Monte Carlo mismatch analysis, extracted-layout simulation, package model, radiation analysis, or measurement is available.  The \SIrange{25}{27}{dB} suppression of the nearest square-wave-LO harmonic products is also modest compared with the image-rejection target and motivates either stronger post-mixer filtering or explicit harmonic-rejection LO synthesis~\cite{kang2018}.

A natural next implementation would therefore couple the frequency plan to a defined antenna/preamplifier, specify a sweep cadence and ADC interface, implement both LO paths, and perform noise, linearity, Monte Carlo, and post-layout analyses.  For a flight-oriented version, electromagnetic compatibility and radiation/thermal qualification would then be added, followed by absolute receiver calibration and measurement.  These steps are required before the architecture can be assessed as a complete astrophysical receiver rather than as the analog channelization core studied here.

\section{Summary and Outlook}
We have studied a frequency-agile Weaver image-reject receiver as the analog channelization stage of a spaceborne low-frequency radio spectrometer.  The architecture accepts signals over \SIrange{0}{50}{MHz}, selects a \SI{3}{MHz} RF slice, and translates it to \SIrange{0}{3}{MHz}.  By pairing the lower and upper halves of the input band as Weaver images and selecting the desired sideband by summation or subtraction, the tunable first LO requires only a \SI{25}{MHz} span; the second LO remains fixed at \SI{26}{MHz}.  Behavioral simulations show that the Weaver topology is better suited than the Hartley alternative to square-wave quadrature drive, while schematic-level end-to-end simulations give \SI{42.89}{dB} IRR at \SI{1}{\degree} phase error and \SI{43.73}{dB} in the all-worst-speed corner.

For astrophysical instrumentation, the main value of the design is not the ideal nominal cancellation but the demonstrated frequency plan: it provides a compact route to tunable sub-band selection without a high-$Q$ RF image filter and reduces the instantaneous bandwidth presented to a downstream digitizer.  The same simulations also expose the principal limitations.  I/Q mismatch sets the useful IRR margin, square-wave LO harmonics require additional attention, and the present work does not yet establish sensitivity, dynamic range, sweep cadence, radiation tolerance, or measured flight performance.  The receiver should therefore be viewed as a validated analog channelizer architecture and a starting point for a complete low-frequency spaceborne radio instrument.

\section*{Acknowledgments}
The receiver design developed in this paper originated in work carried out at the Laboratoire d'\'Etudes Spatiales et d'Instrumentation en Astrophysique (LESIA), Observatoire de Paris/CNRS, and the Laboratoire d'\'Electronique et d'\'Electromagn\'etisme (L2E), Universit\'e Pierre et Marie Curie, and was first documented in the author's Master's thesis~\cite{elsayed2012}.  The present manuscript reorganizes and reanalyzes that material in the context of spaceborne low-frequency radio instrumentation.

\appendix
\setcounter{equation}{0}
\renewcommand{\theequation}{A\arabic{equation}}
\renewcommand{\theHequation}{A.\arabic{equation}}
\section{Detailed Small-Signal Relations}
This appendix collects the small-signal relations used in the circuit analysis.

For the active-load differential stage, the output current is
\begin{equation}
 i_o=i_{D4}-i_{D6}\simeq -g_{m3}(v_+-v_-),
\end{equation}
and the output resistance is $r_{o4}\parallel r_{o6}$, giving (\ref{eq:a2}). For the common-source output stage, the controlled current is $-g_{m7}v_i$ and the load resistance is $r_{o7}\parallel r_{o8}$, giving (\ref{eq:a3}).

For a square-law differential pair,
\begin{align}
 I_{D1}&=K(V_{GS1}-V_t)^2,\\
 I_{D2}&=K(V_{GS2}-V_t)^2.
\end{align}
Writing $v_{id}=V_{GS1}-V_{GS2}$ and $I_{SS}=I_{D1}+I_{D2}$ gives
\begin{equation}
 i_o=I_{D1}-I_{D2}=\sqrt{KI_{SS}}\,v_{id}
 \sqrt{1-\frac{v_{id}^2}{4I_{SS}/K}}.
\end{equation}
For $v_{id}^2\ll4I_{SS}/K$, the square-root term is approximately unity and (\ref{eq:diffpair}) follows.

\end{document}